\documentclass[12pt]{article}
\usepackage{amsmath, graphics,booktabs,setspace}
\usepackage{graphicx,psfrag,epsf}
\usepackage{enumerate}
\usepackage{natbib}
\usepackage{hyperref, hhline}
\usepackage{multicol, multirow}
\usepackage{tikz}
\usetikzlibrary{bayesnet}
\usetikzlibrary{positioning}
\usepackage{adjustbox, bm, amsfonts, dsfont}
\usepackage{placeins, mathtools}
\usepackage{soul} 
\usepackage[title]{appendix}
\usepackage{colortbl}
\newcommand{\blind}{0}

\newcommand{\beginsupplement}{%
        \setcounter{table}{0}
        \renewcommand{\thetable}{S\arabic{table}}%
        \setcounter{figure}{0}
        \renewcommand{\thefigure}{S\arabic{figure}}%
     }

\begin{document}

\def\spacingset#1{\renewcommand{\baselinestretch}%
{#1}\small\normalsize} \spacingset{1}


\if0\blind
{
  \title{\bf A 
  modelling framework for detecting and leveraging node-level information in Bayesian network inference}
  \author{
  Xiaoyue Xi, \\
  Hélène Ruffieux \\
  MRC Biostatistics Unit, University of Cambridge}
  \maketitle
} \fi

\if1\blind
{
  \bigskip
  \bigskip
  \bigskip
  \begin{center}
    {\LARGE\bf Title}
\end{center}
  \medskip
} \fi

\bigskip
\begin{abstract}
Bayesian graphical models are powerful tools to infer complex relationships in high dimension, 
yet are often fraught with computational and statistical challenges. 
If exploited in a principled way, 
the increasing information
collected alongside the data of primary interest 
constitutes an opportunity to mitigate these difficulties by  
guiding  the detection of dependence structures.  
For instance, gene network inference may be informed by the use of publicly available summary statistics on the regulation of genes by genetic variants. Here we present a novel Gaussian  
graphical modelling framework to identify and leverage 
information on the centrality of nodes in conditional independence graphs. 
Specifically, we consider a fully joint hierarchical model to simultaneously infer (i) sparse 
precision matrices 
and (ii) the relevance of node-level information for uncovering the sought-after network structure. We encode such information as candidate auxiliary variables using a spike-and-slab submodel on the propensity of nodes to be hubs, which allows hypothesis-free selection and interpretation of a sparse subset of relevant variables. As efficient exploration of large posterior spaces is needed for real-world applications, we develop a variational expectation conditional maximisation algorithm that scales inference to hundreds of samples, nodes and auxiliary variables.  
We illustrate and exploit the advantages of our approach in simulations and in a gene network study which identifies hub genes involved in biological pathways relevant to immune-mediated diseases. 
\end{abstract}

\noindent%

\spacingset{1.45}
\section{Introduction}
\label{sec:intro}

Undirected graphs are useful tools for expressing relationships between random variables. They are depicted as undirected diagrams where nodes represent variables, and edges represent conditional dependence between nodes, given the remaining nodes in the graph (\emph{partial correlation}). The presence of an edge between two nodes therefore indicates that the corresponding variables are \textit{directly} associated, which provides natural understanding of relationships between variables in many practical applications. Gaussian graphical models provide a framework for estimating such relationships by modelling the variables using a multivariate Gaussian distribution with sparse precision matrix. 
In this setting, a zero entry in the precision matrix is equivalent to a zero partial correlation between the corresponding two Gaussian random variables, that is, the absence of an edge between the nodes. This effectively reduces graph estimation to recovering the support of the precision matrix.

Most existing Gaussian graphical models estimate precision matrices from independent and identically distributed samples of a random vector $(Y_1, \ldots, Y_P)$, by treating the nodes $Y_j$, $j = 1, \ldots, P$, as a priori exchangeable \citep{yuan2007model, friedman2008sparse, wang2012bayesian, li2019expectation}. However, this assumption is often unrealistic, especially when exogenous factors are thought to influence (some of) the nodes and, therefore, the network dependence structure as a whole. In such cases, node-level auxiliary variables may provide additional information on the importance (or ``centrality'' or propensity to have high degree) of each node in the graph structure.

This idea has been employed in the context of regression models to improve variable selection, estimation of regression effects and prediction by encoding predictor-level auxiliary variables. Notably, \cite{van2016better} introduced the concept of ``co-data'' to formalise the task of exploiting external information on predictors, and  \citet{novianti2017better} and \citet{ruffieux2021epispot} proposed further development for high-dimensional regression models, with applications in genomic studies with auxiliary annotation variables (epigenetic marks, probe grouping, conservation status of microRNAs) obtained independently of the outcome. More recently, \cite{li2019expectation} and \cite{bu2021integrating} leveraged edge-wise knowledge, such as the similarity between node attributes, to facilitate edge identification in network estimation. \cite{jewson2022graphical} proposed a graphical lasso framework that exploits auxiliary networks, obtained from external information, to guide inference about the network of primary interest. However, to the best of our knowledge, no graphical modelling approach permits directly accommodating node-level auxiliary variables to aid the detection and interpretation of partial dependence structures. Importantly, such an endeavour differs from directly modifying the prior probabilities of nodes to have high (or low) degree based on some pre-established expert knowledge. Indeed, by \emph{estimating} the effects of candidate node-level variables on the graph structure, their relevance and influence on the nodes are \emph{inferred} from the data in an agnostic fashion. Such a framework may thus improve the estimation of graph structures while enabling the discovery of mechanisms driving these structures. This is particularly relevant for biological networks, which are expected to be \emph{scale-free} \citep{khanin2006scale} – i.e., most nodes have a relatively low degree, while a few nodes, known as ``hubs'', have a high degree – so that node attributes might be leveraged to inform the propensity of nodes to be hubs.

The high dimensionality of graphs, inherent to the quadratic relationship between the number of nodes and the number of parameters, requires enforcing sparsity on the off-diagonal entries of the precision matrix. To achieve this, frequentist methods use regularisation techniques based, e.g., on the lasso penalty \citep{yuan2007model, friedman2008sparse} or the ridge penalty \citep{kramer2009regularized}.  Bayesian methods achieve sparsity using shrinkage priors, typically relying on Markov chain Monte Carlo (MCMC) inference. In particular, Gibbs samplers have been developed for graphical models under double exponential \citep{wang2012bayesian}, spike-and-slab \citep{wang2015scaling} and horseshoe \citep{li2019graphical} priors. However, stochastic search algorithms become computationally inefficient in most real-world problems where the number of nodes exceeds a few tens. To address this, \cite{li2019expectation} subsequently proposed an expectation conditional maximisation (ECM) algorithm as a faster, deterministic alternative to sampling-based inference for a spike-and-slab graphical model. Yet, ensuring scalable inference while allowing for uncertainty quantification for parameters of interest remains difficult in graphical settings.

We tackle the aforementioned challenges by introducing a Bayesian hierarchical framework with contributions at the modelling and inference stages.
First, we propose a fully joint two-level spike-and-slab model that exploits a (possibly) large set of node-level candidate auxiliary variables to estimate the graph structure, while inferring probabilities of informativeness, for each variable, about this structure. Such a formulation is beneficial from both accuracy and interpretability standpoints. Indeed, \emph{inferring} the subset of auxiliary variables associated with the graph structure should improve the estimation of the adjacency matrix and, in turn, that of the precision matrix, by ensuring that only the variables relevant to this estimation are leveraged, with no \emph{ad-hoc} preselection. Moreover, inspecting the global influence of the retained variables on the centrality of nodes may also offer practitioners valuable insights into the exogenous factors in play and their possible role in shaping the network structure. Second, we develop a novel variational Bayes expectation conditional maximisation (VBECM) algorithm that scales comparably to pure ECM algorithms but approximates full posterior distributions rather than point estimates. 

Importantly, our work is concerned with accounting for node-level information, rather than sample-level information to enhance estimation of graphs.
While the latter goal has recently been studied through the formulation of covariate-dependent Gaussian graphical models \citep{ni2022bayesian, zhang2022high}, developing efficient approaches -- both statistically and computationally -- to tackle the former goal is at least as important, given the wealth of external annotations that are now collected alongside datasets of primary interest. As hinted above, this is for instance the case of molecular datasets, for which complementary databases are growing in size and diversity (e.g., about epigenetic mechanisms, gene function and regulation). We will illustrate our framework in a monocyte gene network problem, exploiting summary statistics about the control of genes by genetic variants. Of course, the applicability of our framework extends beyond the field of molecular biology, as the model is generic and free from any domain-specific assumptions. 

This article is organised as follows. 
Section~\ref{sec:data} introduces the monocyte gene expression problem and discusses the potential benefits of encoding gene-level auxiliary information about the genetic control of genes in the network. 
Section~\ref{sec:method} recalls the classical Gaussian graphical model with spike-and-slab prior upon which our approach is based, and details our hierarchical model to select and leverage node-level auxiliary variables.
Section~\ref{sec:inference} describes our VBECM inference algorithm. Section~\ref{sec:simulation} evaluates the statistical and computational performance of our approach in a series of simulations designed to emulate real data settings. Section~\ref{sec:application} applies and exploits our framework on the monocyte data, compares it with a classical Gaussian graphical modelling approach (i.e., with no use of auxiliary information), and discusses potential biological implications of our findings in the context of immune-mediated diseases. Section~\ref{sec:discussion} summarises our work, highlights further application domains and suggests future methodological developments.

\section{Data and motivating example}\label{sec:data}

We introduce a gene expression dataset to discuss the benefits of using relevant annotations for guiding gene network inference and motivate the development of our framework to do so. The data consist of gene expression from 432 European individuals, quantified from CD14$^+$ monocytes using Illumina HumanHT-12 v4 BeadChip arrays \citep{fairfax2012genetics, fairfax2014innate}. Monocytes are myeloid innate immune cells, which play a crucial role in host defence and inflammation by initiating cytokine-mediated response upon microorganism invasion. Therefore, monocytes provide useful clues to investigate disease processes
evaluate potential therapeutic targets and, ultimately, develop novel treatment strategies \citep{ma2019role}.

Conditional independence networks allow pinpointing direct dependencies between genes, and are thus well suited to highlight groups of genes forming signalling pathways, or ``regulatory programs'', that are activated in disease conditions \citep{schafer2005shrinkage}. For instance, Bayesian networks have been used to examine the disruption and conservation of gene pathways during chronic obstructive pulmonary disease \citep{shaddox2018bayesian} and multiple myeloma \citep[a late-stage bone marrow malignancy;][]{ni2022bayesian}. \cite{verdugo2013graphical} analysed monocyte gene levels using co-expression networks and uncovered molecular mechanisms likely underlying atherosclerosis in smokers. The authors estimated indirect relationships (marginal correlation), whereas our focus is on estimating direct relationships (partial correlation).

None of the above studies employed auxiliary annotations and yet, when adequately factored in, the 
growing diversity of 
annotation sources on gene and protein levels offers numerous possibilities for refining the detection of dependence structures. Such annotations include summary statistics on the regulation of genes by genetic variants \citep{ruffieux2020global,kerimov2021compendium}, 
information on biological pathway membership \citep{kanehisa2000kegg, ashburner2000gene}
or scores on gene-level epigenetic activity \citep{budden2016distributed}, to list a few.  

Here we propose to capitalise on information about genetic regulation in monocytes to guide the estimation of network dependence structures. Specifically, major \emph{hotspots} (i.e., genetic variants regulating a large number of nearby and remote genes) have been observed on chromosome 12 in previous monocyte studies \citep{fairfax2014innate, momozawa2018ibd}, and there is evidence that 
genes regulated by a same hotspot are more likely to be functionally related and thus to share edges in the network \citep{van2021trans}. Those in the vicinity of the hotspot also tend to be more tightly controlled by it, and may mediate its effect on other remote genes.
In the network, such mediating genes are typically more connected to other genes and form hubs (``central'' nodes, with high degrees). Thus, accommodating summary statistics on the control of genes by hotspots may aid 
in uncovering gene dependence patterns
including gene hubs. In addition, immune stimulation of CD14$^+$ monocytes, such as through exposure to inflammatory proxies interferon-$\gamma$ (IFN-$\gamma$), tends to trigger enhanced regulatory activity, potentially leading to the formation of additional genetic hotspots \citep{kim2014characterizing, fairfax2014innate} and, as a result, stronger hub patterns in the network.

In this paper, we will construct gene-level auxiliary variables from posterior probabilities of association between genetic variants and genes (our summary statistics) obtained from a genome-wide association study performed with the joint mapping approach \emph{atlasqtl} \citep{ruffieux2020global} using independent genetic and CD14$^+$ monocyte gene expression data \citep{momozawa2018ibd}.
Most of these probabilities are close to zero, reflecting the fact that genetic variants typically control only a few genes, if any.
We will use these summary statistics, in the context of two network estimation problems, namely for genes quantified from resting monocytes as well as IFN-$\gamma$-stimulated monocytes. By encoding information about the genetic regulation of those genes, we hope to not only to improve network inference, but also to suggest genetic variants possibly triggering gene dependence structures, thanks to the hypothesis-free selection of auxiliary variables that is built into our hierarchical spike-and-slab framework. Efficient inference is a prerequisite to applying our approach to the hundreds of genes, samples and tens of candidate auxiliary variables represented in our data. This corresponds to dimensions encountered in many other real-world scenarios, whereby sampling methods, based on MCMC inference, would be computationally prohibitive. The VBECM procedure we will develop is aimed at producing scalable yet accurate inference, with approximations of full posterior distributions.  
We will return to the monocyte problem in Section~\ref{sec:application}.

\section{Methodology}\label{sec:method}

\subsection{The spike-and-slab graphical model}
\label{sec:gss}
We consider a Gaussian graphical model with $N$ observations on $P$ nodes. The data are represented by a matrix $\bm{Y}\in\mathbb{R}^{N\times P}$ whose rows $\bm{y_n}, \ n = 1, \ldots, N$, are independent and identically distributed samples from a multivariate Gaussian distribution, i.e.,
 \begin{eqnarray}\label{eq:gmlik}
\bm{y_1}, \ldots, \bm{y_N} \overset{\text{iid}}{\sim} \mathcal{N}_P(\bm \mu, \bm \Omega^{-1}), \quad \bm{\Omega} \in \mathcal{M}^+,
\end{eqnarray}
where $\mathcal{M}^+$ denotes the set of $P \times P$ symmetric positive definite matrices. Hereafter, we will take $\bm \mu = \bm 0$, without loss of generality.

We assume the precision matrix $\bm{\Omega}$ to be sparse and place a shrinkage prior on its off-diagonal elements and an exponential prior on the diagonal elements. We use a continuous spike-and-slab prior whose formulation allows for the shrinkage of small elements to zero using a mixture of two Gaussian distributions which accounts for the presence or absence of edges,
\begin{eqnarray}\label{eq:cssprior}
p(\bm{\Omega,\delta}\mid \tau, \rho) & \propto & \prod_{i<j} \mathcal{N}\left(\omega_{ij} \mid 0, \frac{\nu^2_{\delta_{ij}}}{\tau}\right) \prod_i \text{Exp}\left(\omega_{ii}\mid \frac{\lambda}{2}\right) \mathds{1} \left(\bm\Omega \in \mathcal{M}^+\right) \prod_{i<j} \text{Bern}\left(\delta_{ij} \mid \rho\right),
\end{eqnarray}
where $\nu_0, \nu_1 > 0$ are set to small and large values respectively ($\nu_0 \ll \nu_1$, see Section~\ref{sec:practical:ssvar}),  $\tau$ is a scaling parameter and $\lambda > 0$ controls the typical size of the diagonal entries. \cite{wang2015scaling} suggests that inference is insensitive to the choice of $\lambda$ since the data typically provide sufficient information for estimating the diagonal; we set $\lambda = 2$ in order to make the diagonal entries a priori equal to one on average. We also assume that each edge is a priori independently included or excluded by modelling the latent binary variable $\delta_{ij}$ as Bernoulli-distributed with unknown probability $\rho$, and place conjugate hyperpriors on $\rho$ and $\tau$, 
\begin{eqnarray}\label{eq:gmrho}
\rho &\sim& \text{Beta}(a_\rho,b_\rho),\\
\tau &\sim& \text{Gamma}(a_\tau, b_\tau),\label{eq:gmtau}
\end{eqnarray}
where $a_\tau = b_\tau = 2$, and $a_\rho  = 1, b_\rho = P$ to induce sparsity \citep{rovckova2014emvs}. 

Alternative shrinkage approaches, such as the horseshoe prior \citep{li2019graphical}, have been proposed for graphical modelling. An advantage of the spike-and-slab formulation is that it permits estimating posterior probabilities of inclusion (PPIs) for the edges which enable direct selection. To perform edge selection and thus estimation of the graph structure, a $0.5$ threshold on the PPIs, resulting in the median probability model \citep{barbieri2004optimal}, may be employed. Alternatively, thresholds corresponding to Bayesian false discovery rates (FDR) can be directly estimated from the PPIs (\citealp{newton2004detecting}; Supplementary Material~1.4).
Under a deterministic inference framework, PPIs are typically well separated \citep{carbonetto2012scalable,rovckova2014emvs}, and different thresholding rules usually produce minor differences in estimations of the graph structure.

We hereafter abbreviate model (\ref{eq:gmlik})--(\ref{eq:gmtau}) as ``GM'' (for vanilla spike-and-slab \underline{g}raphical \underline{m}odel). When setting $\tau = 1$, we recover the model used by \cite{li2019expectation}, which is itself based on the general formulation of \cite{wang2015scaling}.
As discussed in \cite{osborne2022latent}, inferring the scaling factor $\tau$ from the data in continuous spike-and-slab prior specifications allows adaptive learning of the scales and improves edge selection.

\subsection{A framework for leveraging node-level information}
\label{sec:codata}
The GM model (\ref{eq:gmlik})--(\ref{eq:gmtau}) assumes that all nodes in the graph are exchangeable a priori, yet certain nodes may be affected by exogenous factors which make them more likely to have connections. Let $\bm{V}\in\mathbb{R}^{P\times Q}$ be a matrix of $Q$ node-level auxiliary variables, that may be informative on the degree of the $P$ nodes in the network, i.e., its rows $\bm v_i, \ i = 1, \ldots, P$, correspond to ``observations'' (or annotations) on node $Y_i$. As motivated in Sections~\ref{sec:intro} and \ref{sec:data}, we introduce a top-level model hierarchy that lets the propensity of nodes to have high degrees be informed by the $Q$ auxiliary variables via a probit regression on the probability of edge inclusion, 
\begin{eqnarray} \label{eq:codata}
\delta_{ij} \mid \rho_{ij} &\sim& \text{Bern}(\rho_{ij}), \quad 1 \leq i<j \leq P,\\ \nonumber 
\rho_{ij} &=& \Phi\left(\zeta + \sum_{q=1}^Q V_{iq}\beta_q + \sum_{q=1}^Q V_{jq} \beta_q\right),
\end{eqnarray}
where $\Phi(\cdot)$ is the standard normal cumulative distribution function and $\rho_{ij}$ is the edge-specific spike-and-slab probability. Specifically, the inclusion of edge $(i,j)$ depends on the overall network sparsity, controlled by $\zeta$, and on the influence of the annotations for nodes $Y_i$ and $Y_j$. 
Hence, $\beta_q\neq 0$ indicates an effect of variable $V_q$ on the degrees of the nodes in the graph (i.e., on their ``propensity'' to be hubs). 
Other link functions could be employed in place of the probit link, although its use is computationally appealing due to the possibility to employ a data-augmentation formulation that ensures analytical inference updates (Supplementary Material~1.1).
We complete the specification by assuming \begin{eqnarray}\label{eq:zeta}
\zeta &\sim& \mathcal{N}(n_0, t_0^2), 
\end{eqnarray}
and
\begin{eqnarray}\label{eq:codata_normal}
\beta_q \mid \sigma^2 &\sim& N(0, \sigma^2), \qquad q = 1,\ldots, Q, \\ \nonumber
\sigma^{-2} &\sim& \text{Gamma}(a_\sigma,b_\sigma),
\end{eqnarray}
where $a_\sigma = b_\sigma = 2$, 
and $n_0$ and $t_0^2$ are chosen to induce sparsity by matching prior guesses on the expectation and standard deviation for the number of edges in the network (see Section~\ref{sec:practical:overall}). We abbreviate model (\ref{eq:gmlik})--(\ref{eq:cssprior}), (\ref{eq:gmtau})--(\ref{eq:codata_normal}) as ``GMN'', for spike-and-slab \underline{g}raphical \underline{m}odel with \underline{n}ormal prior for the node-level auxiliary variable effects.

\subsection{A framework for selecting node-level information}
\label{sec:codatass}
The top-level regression framework~\eqref{eq:codata}--\eqref{eq:codata_normal} is well suited to settings with a handful of auxiliary variables. Often, however, a large number of candidate auxiliary variables may be available to the practitioner, with the belief that only a few variables (if any) are relevant to the graphical structure of primary interest. We accommodate settings where $Q$ is large by equipping the model with a selection prior that permits leveraging only the variables inferred as being informative, discarding the remaining ones as irrelevant. 
Specifically, we place a spike-and-slab prior on the effects of variables $V_q, \ q = 1, \ldots, Q$, 
\begin{eqnarray} \label{eq:codata_selection}
\beta_q \mid \gamma_q, \sigma^2 &\sim&  \gamma_{q}\mathcal{N}(0,\sigma^2) + (1-\gamma_{q}) \delta(\beta_q) , \quad q=1,\ldots, Q,\nonumber\\ 
\gamma_q\mid o &\sim& \text{Bern}(o), \\ \nonumber
o &\sim & \text{Beta}(a_o, b_o), \\ \nonumber
\sigma^{-2} &\sim& \text{Gamma}(a_\sigma, b_\sigma),
\end{eqnarray}
where $\delta(\cdot)$ is the Dirac delta distribution. The binary latent parameter $\gamma_q$ indicates whether the $q$th variable influences the hub propensity of nodes in the graph, and $o$ is the shared probability for variables to be included. The hyperpriors on $o$ and on the slab variance $\sigma^2$ are conjugate, and we set $a_\sigma=b_\sigma=2$ (as for the GMN model), and $a_o = 1, b_o = Q$ to induce sparsity on auxiliary variables. We refer to model (\ref{eq:gmlik})--(\ref{eq:cssprior}), (\ref{eq:gmtau})--(\ref{eq:zeta}), (\ref{eq:codata_selection}) as ``GMSS'', for \underline{g}raphical \underline{m}odel with \underline{s}pike-and-\underline{s}lab prior for the node-level auxiliary variable coefficients.

As for the bottom-level model on the edges \eqref{eq:gmlik}--\eqref{eq:cssprior}, the spike-and-slab prior formulation \eqref{eq:codata_selection} for the auxiliary variables conveniently yields PPIs which allow directly pinpointing the relevant variables from a potentially large set of candidate auxiliary variables. Hence, thanks to this sparse selection, the GMSS model not only improves the estimation of edges by exploiting variables inferred as relevant to the dependence structures (and only those), but also permits a data-driven discovery of these auxiliary variables. Inspecting the selected variables may be particularly informative to generate hypotheses on the mechanisms underlying the uncovered network dependence structures. 

In summary, inference for the GMSS model yields two sets of spike-and-slab PPIs which will be our main quantities of interest: PPIs based on the binary latent parameters $\{\delta_{ij},\ 1\leq i<j\leq P\}$ for the inclusion of edges, and PPIs based on the binary latent parameters $\{\gamma_{q},\ 1\leq q\leq Q\}$ for the inclusion of auxiliary variables.

\begin{figure}[t!]
  \centering
  \includegraphics[width=0.82\textwidth]{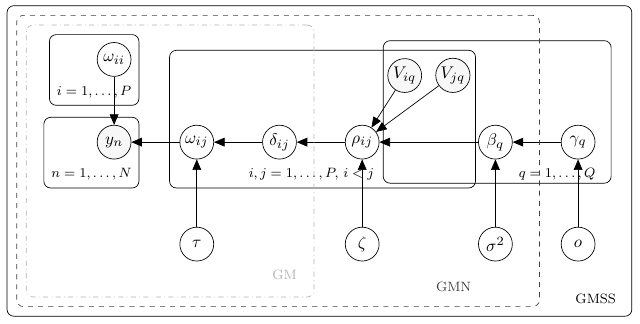}
\caption{\footnotesize \normalfont Graphical representation of the GMSS model (\ref{eq:gmlik})--(\ref{eq:cssprior}), (\ref{eq:gmtau})--(\ref{eq:zeta}), (\ref{eq:codata_selection}). The shaded nodes are observed, and the others are inferred. The grey plates delineate the simpler models detailed in Sections~\ref{sec:gss} and \ref{sec:codata}, i.e., GM: light grey dashed-dotted plate; GMN: dark grey dashed plate.}
\label{fig:graphrep}
\end{figure}

The posterior distribution of the GMSS model is
\begin{eqnarray}\label{eq:posterior}
p(\bm{\Omega}, \bm{\delta},\tau,\zeta, \bm{\beta}, \bm{\gamma}, o, \sigma^2 \mid \bm{Y}) &\propto&  p(\bm{Y} \mid \bm{\Omega})\;\mathds{1}\left( \bm{\Omega} \in \mathcal{M}^+ \right)\prod_{i=1}^P p(\omega_{ii}) \prod_{i<j} \bigg\{ p(\omega_{ij} \mid \delta_{ij}, \tau) p(\delta_{ij} \mid \zeta, \bm{\beta}) \bigg\}  \\ \nonumber
&& \times \,p(\tau) p(\zeta) \prod_{q=1}^Q \bigg\{ p(\beta_q \mid \gamma_q, \sigma^2) p(\gamma_q \mid o) \bigg\} p (o) p(\sigma^2),
\end{eqnarray}
and its graphical representation, as well as those of the GM and GMN models, is provided in Figure~\ref{fig:graphrep}. 

\subsection{Hyperprior elicitation for the network sparsity} 
\label{sec:practical:overall}
In this section, we detail our procedure to elicit the top-level hyperparameters for the prior $\zeta~\sim~\mathcal{N}(n_0, t_0^2)$. While placing hyperpriors on $n_0$ and $t_0^2$ is in principle feasible, we instead choose to limit the model complexity using a hyperparameter elicitation strategy that facilitates interpretation and allows flexibly encoding prior knowledge about the network sparsity. To this end, we repurpose ideas of \cite{bottolo2010evolutionary} and \citet{ruffieux2020global} from the regression setting to the graphical setting. Specifically, we consider the simplifying assumption of a network without the influence of auxiliary variables, i.e., where the probit submodel on the edge spike-and-slab probability $\delta_{ij}$ in \eqref{eq:codata} becomes $\delta_{ij} \mid \zeta \sim \text{Bern}\left\{\Phi(\zeta)\right\}$. Noting that the first two moments for the prior number of edges in the network 
are a function of $n_0$ and $t_0^2$, we then propose to  specify an expectation and a standard deviation for the prior number of edges in the network and solve numerically the system of equations to obtain the corresponding values for $n_0$ and $t_0^2$. Specifying prior numbers of edges in the network has a more intuitive interpretation than directly specifying the top-level hyperparameters $n_0$ and $t_0^2$. 
Detailed derivations and sensitivity analyses are in Supplementary Material~1.2.
These analyses suggest that
inference is robust to the prior expectation of the edge numbers, when the sparse choices are made. The elicitation of the prior standard deviation is even less important, although its influence increases when the prior mean is specified to reflect a graph that is denser than the true graph.

\section{Fast deterministic inference}
\label{sec:inference}
\subsection{Variational expectation conditional maximisation algorithm}
Although block Gibbs sampling has been introduced for spike-and-slab graphical models \citep{wang2015scaling}, MCMC approaches are computationally expensive for high-dimensional graphs due to slow mixing.
To address this, expectation conditional maximisation (ECM) procedures have recently been put forward in the context of graphical models, as faster yet accurate deterministic alternatives to sampling-based approaches \citep{li2019expectation, gan2019bayesian,deshpande2019simultaneous,bai2021spike}.
The ECM algorithm is a generalised expectation maximisation (EM) algorithm, which replaces complicated maximisation steps with several simpler conditional maximisation steps \citep[][see Supplementary Material~1.1.1]
{meng1993maximum}. 
Although computationally convenient, the ECM algorithm provides point estimates only, thus failing to quantify the uncertainty associated with parameter estimates. 
Here we propose a variational Bayes expectation conditional maximisation (VBECM) algorithm,
which couples conditional maximisation (CM) updates for the precision matrix $\bm\Omega$ with approximations of the full posterior distributions of all other model parameters, hereafter gathered in $\bm{\Theta} = (\bm{\delta}, \tau, \zeta, \bm{\beta}, \bm{\gamma}, o, \sigma^2)$.  Specifically, instead of finding point estimates, variational inference aims to
estimate the posterior distribution \eqref{eq:posterior} with an analytical approximate distribution $q(\cdot)$ that minimises the reverse Kullback–Leibler divergence with the true posterior,
\begin{eqnarray*}
\text{KL}(q \mid\mid p) = - \mathbb{E}_{q(\bm{\Omega}, \bm\Theta)} \log \Bigg\{ \frac{p(\bm{\Omega}, \bm{\Theta} \mid \bm{Y})}{q(\bm{\Omega}, \bm{\Theta})}\Bigg\},
\end{eqnarray*}
where $\mathbb{E}_{q(\bm{\Omega}, \bm\Theta)}(\cdot)$ is the expectation with respect to the variational distribution $q(\bm{\Omega}, \bm\Theta)$.
This is equivalent to maximising the following a lower bound on the marginal log-likelihood,
\begin{eqnarray}\label{eq:elbo}
\mathcal{L}(q) & = &\mathbb{E}_{q(\bm{\Omega}, \bm\Theta)} \log p(\bm{Y},\bm{\Omega}, \bm{\Theta}) - \mathbb{E}_{q(\bm{\Omega}, \bm\Theta)} \log  q(\bm{\Omega}, \bm{\Theta}),
\end{eqnarray}
which is also known as the \emph{\underline{e}vidence \underline{l}ower \underline{bo}und} \citep[ELBO;][]{bishop2006pattern}. Since the ELBO does not involve the expression of the marginal likelihood, it can conveniently be used as an objective function.

To find the optimal variational distribution, we rely on a \emph{mean-field} approximation that assumes independence for a partition of the model parameters as follows,
\begin{eqnarray}\label{eq:mf}
q(\bm{\Omega}, \bm{\Theta})
& = & q(\bm{\Omega}) \prod_{i<j} q(\delta_{ij}) q(\tau) q(\zeta) \prod_{q=1}^Q q(\beta_q, \gamma_q) q(o) q(\sigma^2).
\end{eqnarray}
This factorisation is based on the following considerations. First, since $\bm{\Omega}$ is positive definite, its entries cannot be treated independently. We thus model it using a multivariate variational factor.
The spike-and-slab parameters for each auxiliary variable $V_q$, $q = 1, \ldots, Q$, are also modelled jointly based on the factorisation
\begin{eqnarray}\label{eq_struct_ss}
q(\beta_q, \gamma_q) = q(\beta_q\mid \gamma_q) q(\gamma_q), 
\end{eqnarray}
which retains the mixture structure and thus posterior multimodality. The remaining independence assumptions enable closed-form updates; their possible influence on inference will be assessed in simulations through evaluation of the edge and variable selection performance (Section~\ref{sec:sim:accuracy}). 

We perform the optimisation with a coordinate ascent algorithm that iteratively optimises each factor of the mean-field approximation~\eqref{eq:mf} while keeping the other factors fixed. The
variational distribution for $\bm{\Omega}$ is intractable,  
so we 
obtain a conditional maximisation instead.
Specifically, we reframe \cite{wang2015scaling}'s block Gibbs sampling procedure as a CM step in an ECM procedure, which is equivalent to considering the variational distribution of $\bm \Omega$ as being 
 a point mass.
The full derivations are provided in Supplementary Material~1.1.2.

\subsection{Parallel grid search procedure for spike-and-slab variances}
\label{sec:practical:ssvar}
The GMSS model relies on two distinct spike-and-slab formulations: a continuous formulation for the edge effects, where the spike component is modelled using a peaked Gaussian distribution with variance $\nu_0^2$ smaller than the slab variance $\nu_1^2$, and a discrete formulation for the auxiliary variable effects, where the spike component is modelled using a Dirac delta point mass at zero. Continuous spike-and-slab priors have been used to investigate theoretical properties of mixture priors \citep[][]{george1993variable, ishwaran2005spike, narisetty2014bayesian}, but tend to be less commonly used than their discrete counterpart, partly because they require specifying two variance parameters ($\nu_0^2$ and $\nu_1^2$) instead of one only ($\nu_1^2$) for discrete spike-and-slab priors. However the original spike-and-slab graphical model by \cite{wang2015scaling} and its subsequent variants \citep[e.g., ][]{li2019expectation, li2019bayesian} use the continuous spike-and-slab to model edge effects as this allows obtaining Gibbs samplers or ECM algorithms that maintain the positive definiteness constraint on the precision matrix at each iteration (provided a positive definite matrix is used to initialise $\bm \Omega$). Here we rely on the same updating strategy (see Supplementary Material~1.1.2)
and thus use a hybrid formulation with a continuous spike-and-slab prior to model edge effects and a discrete spike-and-slab prior to model auxiliary variables effects. 

Appropriately eliciting the variances $\nu_0^2$ and $\nu_1^2$ in the continuous spike-and-slab formulation \eqref{eq:cssprior} is crucial as, along with the hyperparameter setting for $\zeta$ (Section~\ref{sec:practical:overall}), their values determine the level of regularisation. In principle, 
both variances could be inferred simultaneously,
however, this tends to yield 
degenerate solutions under empirical Bayes settings when the problem is sparse \citep{scott2010bayes, van2019learning}. 
To bypass this issue, proposals have been made to explore a series of spike variances while keeping the slab variance fixed
\citep{rovckova2014emvs, rovckova2018spike, li2019expectation}. 
Several routes can be explored to
set 
the spike variance based on  
a prior guess of the network sparsity \citep{osborne2022latent}, cross-validation procedures \citep{li2019expectation}, or model selection criteria such as the Akaike information criterion \citep[AIC;][]{akaike1998information, langfelder2007eigengene}, the Bayesian information criterion \citep[BIC;][]{schwarz1978estimating, jewson2022graphical} and the extended Bayesian information criterion  \citep[EBIC;][]{chen2008extended}. Alternatively, it has been proposed to choose the spike variance by inspecting the stabilisation of precision matrix estimates when conducting sequential runs of the algorithm whereby the variance is decreased progressively. For each run, parameters are initialised using the ``optimal'' values estimated at the previous run, thus resulting in a \emph{dynamic posterior exploration} \citep{rovckova2018spike, bai2021spike, li2019graphical}.

In practice, the sparsity is unknown and rough estimates are typically difficult to obtain without conducting full network inference, so selection based on sparsity is impractical. Cross-validation procedures have no guarantee for model selection consistency \citep{fan2013tuning} and are computationally prohibitive in realistic graphical settings. 
Dynamic posterior exploration has gained popularity in recent years, but guarantees of convergence to the optimal solution have yet to be established.

In this paper, we propose a parallel grid-search procedure based on model selection criteria.
Specifically, we set $\nu_1 = 100$ and consider a series of candidate values for $\nu_0$ from grid ranging from $10^{-2}$ to $1$. We then run the VBECM algorithm for the different choices of $\nu_0$ and select the value corresponding to the run with the lowest AIC. Our simulations show that the selected value usually corresponds to settings with the highest edge selection performance and good auxiliary variable selection performance and other model selection criteria produce similar results (Supplementary Material~1.3).
Moreover, the selected $\nu_0$ is typically small, resulting in a small spike variance and thus small estimated precision matrix entries for edges inferred as ``absent''. Importantly, the absence of sequential updates allows us to launch all the runs in parallel, which results in a highly efficient search. In particular, if the number of cores equals or exceeds the number of values in the grid, the runtime is dominated by the most computationally intensive run across the grid.

Our approach is implemented in the publicly available R package 
\texttt{navigm} (\underline{n}ode-level \underline{a}uxiliary \underline{v}ariables for \underline{i}mproved \underline{g}raphical \underline{m}odel inference).

\section{Simulations}
\label{sec:simulation}

\subsection{Data generation and simulation set-up}\label{sec:simulation:setup}
The numerical experiments presented below are meant to (i) assess the performance of our approach for estimating edges (adjacency matrix) and identifying ``active'' variables, i.e., node-level variables that are informative on the degrees or ``centrality'' of nodes, and therefore on the dependence structures in the network;
(ii) benchmark it against state-of-the-art graphical modelling approaches on synthetic data designed to emulate real settings. 
We simulate networks with $N$ samples and $P$ nodes as follows. 
First, we generate $Q$ candidate auxiliary variables from a right-skewed beta distribution with parameters $0.05$ and $0.2$, resulting in entries being mostly close to 0 with a few close to 1, and we gather them in a $P \times Q$ matrix $\bm{V}$.  We further randomly select $Q_0 \leq Q$ active variables, and simulate their effects from a log-normal distribution with mean 0.5 and standard deviation 0.1; these effects are positive as we are particularly interested in emulating problems where features are associated with the presence of hubs, as motivated in Sections~\ref{sec:intro} and~\ref{sec:data}. Although negative effects of auxiliary variables may not be a primary concern in practice, our modelling framework can also detect such effects, as it enforces no restriction on the sign of effects. For completeness, we describe simulations in Supplementary Material~2.3
that comprise negative effects and a combination of positive and negative effects.
 
Unless stated otherwise, we choose $\zeta$ such that the network sparsity is $\approx 3\%$. We then construct a binary adjacency matrix $\bm{A}$ for the graph skeleton: we establish its $(i,j)$th entry by thresholding the probability parameter $\rho_{ij}$ in (\ref{eq:codata}) at $0.5$ (median probability model), and 
set to $1$ an additional percentage of zero entries in the
upper triangular part 
 of $\bm A$ to include edges that are not induced by auxiliary variables (hereafter set to $10\%$, unless otherwise specified and referred to as ``noise level''). Next, we generate the precision matrix $\bm{\Omega}$ based on the structure of the adjacency matrix as follows \citep{tan2014learning},
\begin{eqnarray*}
E_{ij} &\sim& 
  \begin{cases} 
      0, & A_{ij} = 0, \\
      \text{Unif}(-0.75,-0.25) \cup \text{Unif}(0.25,0.75), & A_{ij} \neq 0,
   \end{cases}
   \\
\bm{\bar{E}} &=& \frac{1}{2}(\bm{E} + \bm{E}^T ),\\
\bm{\Omega} &=& \bm{\bar{E}}  + (0.1 - \lambda_E^\text{min})\bm{I}_P,
\end{eqnarray*}
where $\bm{I}_P$ refers to the $P \times P$ identity matrix, and $\lambda_E^\text{min}$ represents the smallest eigenvalue of $\bm{E}$. This construction guarantees that the precision matrix $\bm{\Omega}$ is symmetric positive definite and matches the structure of the adjacency matrix $\bm{A}$.  
Finally, we simulate $N$ samples independently from the multivariate normal distribution $\mathcal{N}_P(\bm{0}, \bm{\Omega^{-1}})$. 
In each experiment, we generate $100$ data replicates and summarise the performance across all the replicates. 
R~code for reproducing all simulation results presented in the subsequent sections is available at \url{https://github.com/XiaoyueXI/navigm_addendum}.

\subsection{Edge and auxiliary-variable selection performance}\label{sec:sim:accuracy}

In this section, we evaluate the edge-selection performance of our modelling framework when encoding external information on the network structure, as well as its ability to identify the auxiliary variables relevant to this structure. 
We discuss two scenarios: a problem with a small number of auxiliary variables, where no selection is needed, and a problem with a large number of candidate variables of which only a subset is active, thus benefiting from sparse selection.  

For the first scenario, we simulate data with $N=200$ samples and $P=100$ nodes whose hub structure is influenced by 
three variables, i.e., $Q=3$ and $Q_0=3$. We compare the performance of the vanilla GM model, which does not incorporate any auxiliary information, with that of the GMN model, which represents the effects of the auxiliary variables using a normal prior (see Figure~\ref{fig:graphrep}). 
To ensure comparability between the two models, we replace the GM model's beta prior specification in \eqref{eq:gmrho} with a normal prior 
via a probit submodel, namely, $\rho = \Phi(\zeta)$ (Supplementary Material~
2.1);
we hereafter refer to this modified model as ``GM$^*$''.

\begin{figure}[ht!]
\centering
\includegraphics[width=0.75\textwidth]{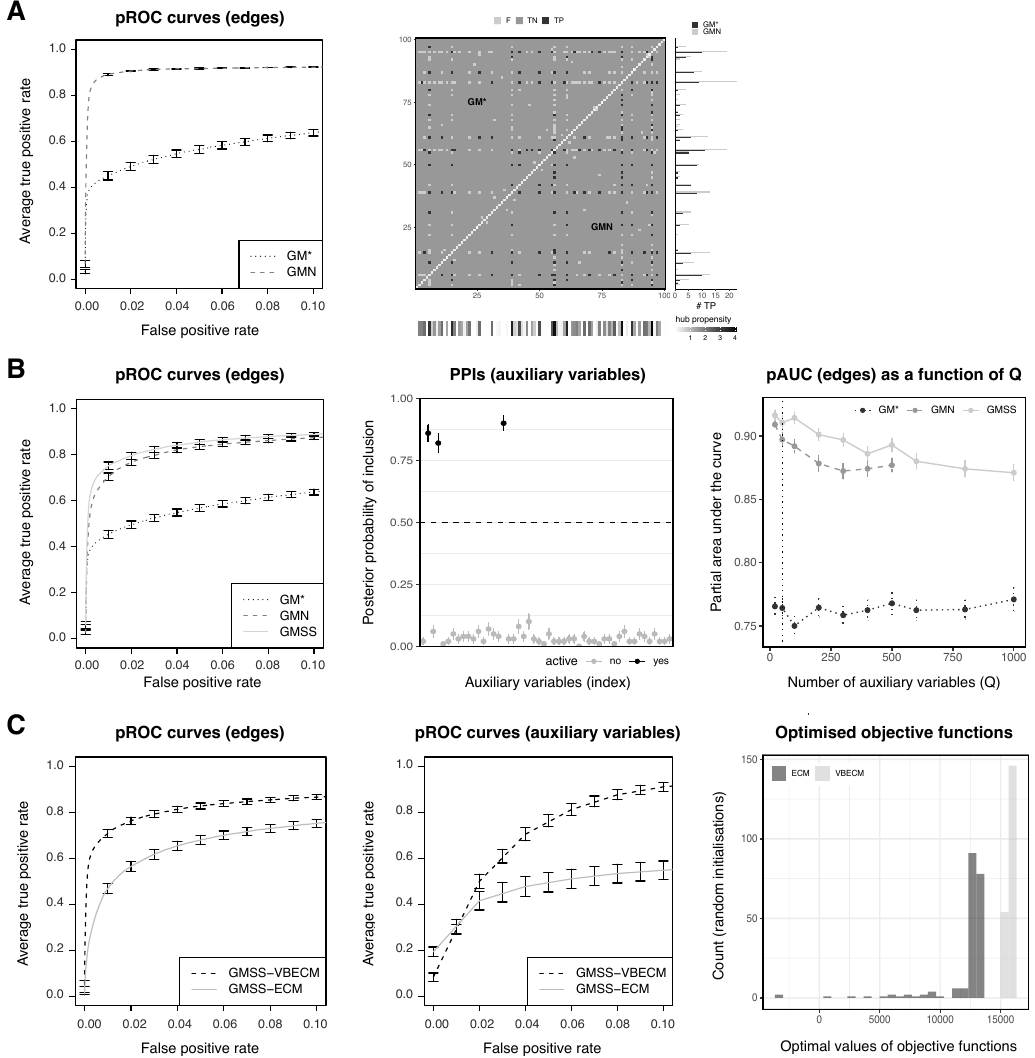}
\caption{\footnotesize \linespread{1.0}\selectfont Performance for a problem with $N = 200$ samples, $P = 100$ nodes and 3 active auxiliary variables (100 replicates) under different scenarios. (A): Problem with a small number of candidate auxiliary variables, i.e., $Q=3$; (B): Problem with a large number of candidate auxiliary variables, i.e., $Q=50$; (C): Comparison of VBECM and ECM inference for the GMSS model, on a problem with $Q=50$ variables. (A–C, left): Average pROC curves, with standard error bars for edge selection. (A, right): Heatmap of the graph recovery performance for the first replicate. Each entry corresponds to a pair of nodes (presence/absence of edges between them), with ``F'' (false positives or false negatives, light grey), ``TN'' (true negatives, dark grey) and ``TP'' (true positives, black), using GM$^*$ (top-left triangle) and GMN (bottom-right triangle). The contribution of auxiliary variables in influencing the node degrees (or ``hub propensity''), i.e., $\bm{v_{i} \mathbb{E}_{q(\bm \Omega, \Theta)}(\beta) }$ for node $i$, is shown in the bottom margin. The total number of true positive edges for each node is compared between GM$^*$ (black) and GMN (grey) in the right margin. (B, middle): Average PPIs of the candidate auxiliary variables, estimated with GMSS, with standard error bars. The variables simulated as active are in black, and the others in grey. 
(B, right): Average pAUC for edge-selection as a function of $Q$, along with standard errors. The line for GMN is truncated as the method could not converge within 1.5 day on an Intel Xeon CPU, 2.60 GHz machine. The vertical black dotted line represents the edge selection performance in the reference scenario. (C, middle): Average pROC curves for auxiliary variable selection using GMSS. (C, right): Optimal values of the objective function reached in 200 runs of the GMSS-ECM ($Q$ function) and GMSS-VBECM (ELBO) algorithms using different random starts for the first replicate.}
\label{fig:small_ref_inference}
\end{figure}

We assess performance using average partial receiver operating characteristic (pROC) curves, as follows: for each of $100$ data replicates, we obtain a pROC curve by applying varying thresholds to the inferred posterior probabilities of inclusion (of edges or auxiliary variables), for false positive rates between $0$ and $0.1$. We then construct an average curve across replicates, with standard error bars. As expected, Figure~\ref{fig:small_ref_inference}A indicates that the encoding of the three variables aids the recovery of edges in the network. Inspecting the reconstructed graph for a given replicate shows that inference using the GMN model results in a larger number of true positives compared to the GM$^*$ model. 
This example, where all candidate variables contribute to the node dependence structure, serves as an ``easy'' setting to check that the model indeed improves network inference when relevant auxiliary information is used. 

We next consider the more challenging setting where a large number of candidate auxiliary variables, $Q = 50$, are measured, yet only sparsely related with the network structure, i.e., with zero effect for most variables, except for $Q_0 = 3$ 
active variables. We refer to this setting as our 
``reference'' data generation scenario. With its top-level spike-and-slab formulation, the GMSS approach accounts for the sparse nature of the auxiliary variable effects, while also providing intuitively appealing PPIs for each variable, which are unavailable from the GMN approach. 

Figure~\ref{fig:small_ref_inference}B compares GMSS, GMN and GM$^*$, and again shows an improved edge selection performance when the auxiliary variables are leveraged, with GMSS and GMN 
(see also Supplementary Material~
2.2).
In addition, while the estimation of the \emph{graph structure} is the primary focus of our work, the \emph{precision matrix estimates} also indirectly benefit from the encoding of auxiliary data via the GMSS spike-and-slab binary latent variables. Indeed, the average mean absolute error of precision matrix estimates for true edges reported by either method reduces from $0.49$ 
(s.e. $0.00$) using GM$^*$ to $0.22$ (s.e. $0.01$) using GMSS. Figure~\ref{fig:small_ref_inference}B also shows a modest advantage for GMSS compared to GMN, likely due to the validity of sparsity assumption on the auxiliary variable effects and the accurate recovery of these effects. Indeed, 
the three variables relevant to the graph structure are correctly identified by GMSS: 
the average PPIs for these variables exceed $0.5$, while those for the remaining variables, simulated as ``inactive'', are below $0.25$. 
From now on, our primary intent in discussing GMN is
to contrast it with GMSS and underscore the possible the statistical and computational advantages of the latter method which is tailored to large $Q$ data settings.

\begin{table}[!t]
\centering\resizebox{\textwidth}{!}{
\begin{tabular}{|c|c|c|c|c|c|c|cccc|} 
\hline
& &  & & & & & \multicolumn{3}{c|}{Edge selection} & \multicolumn{1}{c|}{Variable selection} \\
\cline{8-11}
 & Sparsity & Noise   & $Q_0$ & $Q$ & $N$ & $P$ & \multicolumn{1}{c|}{GM$^*$} & \multicolumn{1}{c|}{GMN} &  \multicolumn{2}{c|}{GMSS\hspace{1cm}\mbox{}} \\
\hline
1. &
\cellcolor{gray!25} &  
\cellcolor{gray!25}
&  
\cellcolor{gray!25}
& \cellcolor{gray!25} & \cellcolor{gray!25} & \cellcolor{gray!25} 100 & 0.76 (0.01)  & \textbf{0.90 (0.01)}  & \textbf{0.91 (0.01)}  & 0.90 (0.01) \\
\hhline{~~~~~~-----}
2. &\cellcolor{gray!25}& \cellcolor{gray!25}& \cellcolor{gray!25}&  \cellcolor{gray!25} &\multirow{-2}{*}{\cellcolor{gray!25}200} & 50 & 0.89 (0.01)  & \textbf{0.94 (0.00)}  & \textbf{0.93 (0.01)}  & 0.84 (0.01)  \\
\hhline{~~~~~------}
3. &\cellcolor{gray!25}& \cellcolor{gray!25}& \cellcolor{gray!25}& \cellcolor{gray!25} & \multirow{2}{*}{100} & \cellcolor{gray!25}100 & 0.72 (0.01)  & \textbf{0.84 (0.01)}  & \textbf{0.83 (0.01)}  & 0.84 (0.01) \\
\hhline{~~~~~~-----}
4. &\cellcolor{gray!25}& \cellcolor{gray!25}& \cellcolor{gray!25}& \multirow{-4}{*}{\cellcolor{gray!25}50} & &
50 & 0.83 (0.01)  & \textbf{0.91 (0.01)}  & \textbf{0.90 (0.01)}  & 0.80 (0.01) \\
\hhline{~~~~-------}
5. &\cellcolor{gray!25}& \cellcolor{gray!25} & \cellcolor{gray!25} & 20 & \cellcolor{gray!25}  &  \cellcolor{gray!25} & 0.77 (0.01)  & 0.91 (0.00)  & \textbf{0.92 (0.00)}  & 0.87 (0.01) \\
\hhline{~~~~-~~----}
6. &\cellcolor{gray!25}&\cellcolor{gray!25} & \multirow{-6}{*}{\cellcolor{gray!25}3}& 100 & \cellcolor{gray!25} &\cellcolor{gray!25} & 0.75 (0.01)  & 0.89 (0.01)  & \textbf{0.91 (0.00)}  & 0.92 (0.01)  \\
\hhline{~~~--~~----}
7. &\cellcolor{gray!25}& \cellcolor{gray!25}&  1 & \cellcolor{gray!25} & \cellcolor{gray!25} & \cellcolor{gray!25} &   0.74 (0.01)  & 0.95 (0.00)  & \textbf{0.96 (0.00)}  & 0.99 (0.00) \\
\hhline{~~~-~~~----}
8. &\cellcolor{gray!25}& \multirow{-8}{*}{\cellcolor{gray!25}10\%}& 5 & \cellcolor{gray!25} & \cellcolor{gray!25} & \cellcolor{gray!25} & 0.76 (0.01)  & \textbf{0.89 (0.01)}  & \textbf{0.88 (0.01)}  & 0.82 (0.01)  \\
\hhline{~~--~~~----}
9. &\cellcolor{gray!25}& 20\% & \cellcolor{gray!25} &  \cellcolor{gray!25} & \cellcolor{gray!25} & \cellcolor{gray!25} &
0.77 (0.01)  & \textbf{0.88 (0.01)}  & \textbf{0.88 (0.01)}  & 0.91 (0.01)  \\
\hhline{~~-~~~~----}
10. &\multirow{-10}{*}{\cellcolor{gray!25}3\%} & 30\% & \cellcolor{gray!25} & \cellcolor{gray!25} & \cellcolor{gray!25} & \cellcolor{gray!25} & 0.77 (0.01)  & 0.84 (0.00)  & \textbf{0.86 (0.00)}  & 0.90 (0.01)  \\
\hhline{~--~~~~----}
11. &1\% &  \cellcolor{gray!25} & \cellcolor{gray!25}  & \cellcolor{gray!25} & \cellcolor{gray!25} & \cellcolor{gray!25} & 
0.87 (0.01)  & \textbf{0.95 (0.00)}  & 0.94 (0.00)  & 0.85 (0.01) \\
\hhline{~-~~~~~----}
12. &8.5\% & \multirow{-2}{*}{\cellcolor{gray!25}10\%} & \multirow{-4}{*}{\cellcolor{gray!25}3} & \multirow{-6}{*}{\cellcolor{gray!25}50} & \multirow{-6}{*}{\cellcolor{gray!25}200}& \multirow{-8}{*}{\cellcolor{gray!25}100} &  0.66 (0.00)  & 0.83 (0.01)  & \textbf{0.89 (0.01)}  & 0.93 (0.01)  \\
\hline
\end{tabular}
}
\caption{\footnotesize \normalfont Performance for a grid of $12$ data generation scenarios, using the GM$^*$, GMN and GMSS models, showing the mean and standard error (in parentheses) of pAUCs for edge selection (for GM$^*$, GMN and GMSS) and auxiliary variable selection (for GMSS) based on 100 data replicates. Scenario 1 (first row) corresponds to the ``reference'' simulation presented in Figure~\ref{fig:small_ref_inference}B, and shaded cells indicate the same settings as this scenario. For each scenario, the average pAUCs within one standard error of the highest average pAUC are highlighted in bold.
}
\label{tab:sim:sensitivity}
\end{table}

Table~\ref{tab:sim:sensitivity} generalises these observations to different data generation scenarios, namely, for a grid of average sparsity and noise levels, and numbers of candidate auxiliary variables $Q$, active variables $Q_0$, samples $N$ and nodes $P$. It shows the average standardised partial areas under the curve (pAUCs) for the edge and auxiliary variable selection performance. As expected, irrespectively of the modelling approaches (GM$^*$, GMN, GMSS), edge estimation is more reliable when the network sizes are small (compare scenarios 1 with 2, and 3 with 4) and sample sizes are large (compare scenarios 1 with 3, and 2 with 4). Moreover, for given network sizes and sample sizes, estimation typically gets more challenging as (i)~the graph gets denser (compare, e.g., scenarios 11 with 12), (ii)~the number of active auxiliary variables gets larger (compare, e.g., scenarios 7 with 8), or (iii)~the noise level (proportion of edges not influenced by auxiliary variables) gets larger (compare scenarios 1, 9 and 10). 

We also find that GMN and GMSS always outperform GM$^*$ in terms of edge selection. Moreover, as for the previous simulation example (Figure~\ref{fig:small_ref_inference}B), the sparse selection of auxiliary variables induced by the top-level spike-and-slab submodel of GMSS results in comparable or improved graph recovery compared to GMN. Figure~\ref{fig:small_ref_inference}B also presents this comparison as a function of $Q$; it confirms a modest yet consistent outperformance of GMSS over GMN. Importantly, GMN becomes computationally intractable as $Q$ increases,  failing to converge within 1.5 days as $Q>500$.
In general, GMSS can save a factor of up to $3$ 
in computational time compared to GMN 
(Supplementary Material Table~2),
likely due to the validity of the sparsity assumption again: indeed the spike-and-slab prior (GMSS) permits a more efficient exploration of the posterior space compared to the ``non-sparse slab-only'' normal prior (GMN). Finally, Table~\ref{tab:sim:sensitivity} shows that the GMSS average pAUCs for the auxiliary variable selection based on the estimated PPIs range from $0.80$ to $0.99$, which suggests a very good recovery of the variables relevant to the graph structure; again, such PPIs constitute directly interpretable posterior quantities which confers a clear advantage to the GMSS approach when $Q$ exceeds a handful of variables.

\subsection{Null scenario and robustness to model misspecification}\label{sec:simulation:null}
We next discuss two simulation scenarios which depart from data settings for which GMSS is primarily designed for. We first consider a scenario where the graph dependence structure is not influenced by any auxiliary information, i.e., the top-level auxiliary variable model is a null model ($Q_0=0$). Our goal is to evaluate the behaviour of our approach when $Q=50$ irrelevant variables are used as candidate auxiliary variables. 
GMSS correctly discards all variables as irrelevant to the graph structure, as their PPIs are all very low $<0.1$. This suggests that large panels of candidate variables, whose relevance for the underlying graph is unclear, can be safely supplied to the GMSS approach: the spike-and-slab prior formulation not only allows pinpointing the relevant variables but also ensures that the irrelevant variables are correctly dismissed as ``noise''. 
In contrast, GMN, which relies on a ``slab-only'' Gaussian prior, cannot handle a large number of (irrelevant) variables. It produces unstable inferences, with highly variable estimates across replicates (Supplementary Material~2.4).
Such insufficient regularisation of the auxiliary variable effects translates into a significantly lower edge selection performance (average pAUC: 0.55, s.e. $< 0.01$) compared to GMSS (average pAUC: 0.70, s.e. $< 0.01$). Our explorations further indicate that the GMN algorithm can be sensitive to parameter initialisations, especially in weakly informative data settings.

We next evaluate the robustness of GMSS to model misspecification in a simulation setting where the \emph{similarity between attributes of pairs of nodes} influences the presence or absence of edges between the nodes, but not the \emph{values} of the node attributes (encoded as auxiliary variables) themselves. For instance, brain connectivity networks describe the connectivity between regions of interest (ROIs) within individual brains. They are typically modelled by Gaussian graphical models based on functional magnetic resonance imaging (fMRI) signals at each of the ROIs for a large number of images obtained during a scanning session \citep{higgins2018integrative}. The distance between these ROIs may influence their connectivity \citep{bu2021integrating}.
We consider a simulation study assuming a misspecified edge regression model to mimic the aforementioned example. For each of $100$ replicates, we simulate $50$ variables of which $2$ are relevant in the sense that edges are more likely to be present between nodes with similar values of these two variables. We generate datasets with $N=200$ samples and $P=100$ nodes under such a similarity-based edge model and use GMSS to estimate the graph and effects of auxiliary variables. 
In this setting, GMSS exhibits the desirable behaviour of discarding all auxiliary variables as irrelevant (PPIs $< 0.1$) to the hub propensity of the nodes, akin to the null model scenario case (see Supplementary Material~2.5
for details).

\subsection{Comparison with existing inference approaches}\label{sec:sim:comparison}

As motivated earlier, no graphical modelling approach exploiting auxiliary node-level variables exists, to our knowledge. It is nevertheless important to assess the accuracy and robustness of our VBECM algorithm by comparing it to other inference algorithms. 
To this end, we first consider the EMGS approach proposed by \cite{li2019expectation}, which is based on a model similar to the GM model presented in Section~\ref{sec:gss} (see also a graphical representation of Figure~\ref{fig:graphrep}), with the following differences: (i) the GM model has added flexibility with a scale parameter $\tau$ in the edge-level spike-and-slab prior (fixed to $1$ for EMGS), and (ii) the GM model parameters are inferred using variational inference (rather than pure ECM in EMGS). 

Here we focus on evaluating the benefits of (ii), that is, of full posterior distribution estimation by variational inference. To this end, we compare our VBECM implementation for the GM model (hereafter referred to as ``GM-VBECM''), with a pure ECM implementation of the same model (``GM-ECM'').
We generate $100$ datasets under the vanilla scenario where the graph structure does not depend on auxiliary information, for a problem with $N = 200$ samples and $P = 100$ nodes. The edge-selection performance of the GM-VBECM and GM-ECM runs is almost the same, based on the same hyperparameter and initialisation choices to ensure fair comparisons. 
 The computational time is comparable for the two inference algorithms, both of which take less than 2 seconds on average.

We next compare VBECM and ECM inference for the GMSS model, based on data where the graph structure is influenced by auxiliary variables. We use again the ``reference'' data generation scenario, with $N = 200$ samples, $P= 100$ nodes and $Q=50$ auxiliary variables, of which $Q_0 = 3$ contribute to the node degrees. Figure~\ref{fig:small_ref_inference}C shows the average pROC curves for edge selection and auxiliary variable selection. Here, the benefits of approximating the full posterior distributions with variational inference are clear. The improved performance is likely largely imputable to the ability of the VBECM algorithm to capture parameter uncertainty – notably thanks to the expressive joint variational approximation (\ref{eq_struct_ss}) for $\beta_q$ and $\gamma_q$ – and, as a result, to be less prone to entrapment in local modes, unlike the ECM algorithm. This is suggested by the large variability of optimal objective functions
obtained from multiple restarts of the ECM algorithm applied to the first data replicate; in contrast, the VBECM algorithm consistently reaches similar values for the optimised ELBO (Figure~\ref{fig:small_ref_inference}C and Supplementary Material~2.6).
A comparison of ECM and VBECM for the effects of the auxiliary variables further show that the variational posterior means are more accurate than the ECM point estimates. Additionally, unlike GMSS-ECM, GMSS-VBECM produces credible intervals, and these cover the true value in most simulation replicates, despite the well-known tendency of variational inference to underestimate posterior variances 
(Supplementary Material~2.7).

Finally, unlike variational updates which can be obtained for model components with discrete point mass distributions, the ECM algorithm necessitates the use of a continuous spike-and-slab prior for the top-level auxiliary variable effects, as it requires taking derivatives of the objective function. This implies the use of an additional grid search for the spike variance, as described for the bottom-level edge effects (Section~\ref{sec:practical:ssvar}). This requirement may also partly contribute to the performance gap observed in Figure~\ref{fig:small_ref_inference}C, and impacts computational efficiency substantially; the average runtime for the example of Figure~\ref{fig:small_ref_inference}C is 
$\approx 6$ minutes for GMSS-ECM and only $40$ 
seconds for GMSS-VBECM.

\subsection{Runtime profiling}\label{sec:sim:runtime}

\begin{figure}[!t]
\centering    
\includegraphics[width=0.9\textwidth]{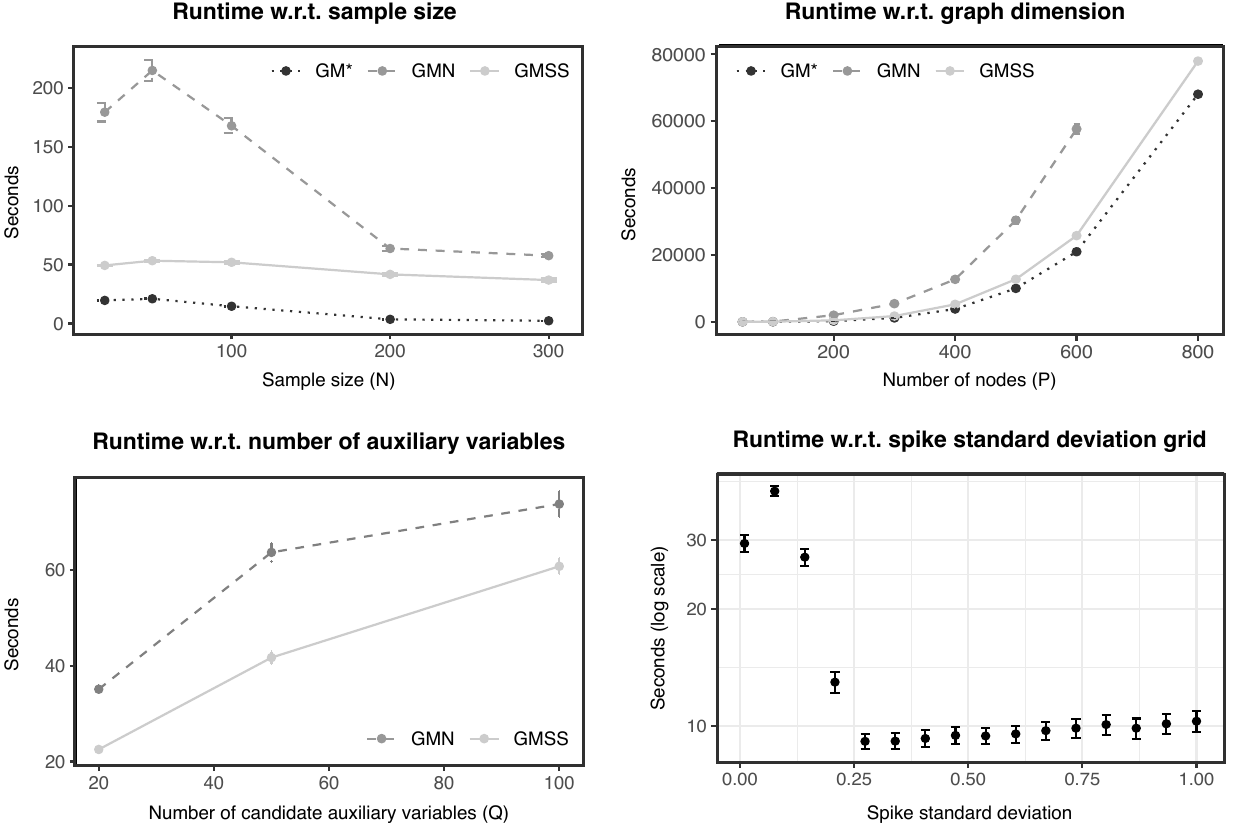}
\caption{\footnotesize \normalfont \linespread{1.0}\selectfont Runtime profiles in seconds on an Intel Xeon CPU, 2.60 GHz. Average time until convergence in seconds against sample sizes (top left), network sizes (top right) and the number of auxiliary variables (bottom left), with standard error bars based on $100$ replicates. Bottom right: runtime breakdown for the ``reference'' scenario GMSS, with 
$N=200, P=100, Q=50$, for a series of spike standard deviations on the $x$-axis. The runtimes reported in all the panels except the bottom right panel include the grid search procedure. }
\label{fig:runtime}
\end{figure}

This section presents a systematic assessment of the computational efficiency of our framework. We again consider a grid of scenarios, based on variations of the ``reference'' scenario used in previous experiments. Specifically, Figure~\ref{fig:runtime} shows the runtimes obtained with GM$^*$, GMN, and GMSS when the sample sizes $N$, 
network sizes $P$,
or number of auxiliary variables $Q$ is varied. In all cases, the GMN approach requires the largest number of iterations to converge, making it the most computationally demanding approach, while the GM$^*$ approach is the fastest as it relies on a simpler model with no top-level hierarchy for auxiliary variables. All methods struggle to converge for very small sample sizes (although they all do converge, eventually) and the runtime is not increased as more information becomes available with larger sample sizes ($N\leq 300$). However, because of the quadratic relationship between the number of nodes and the number of parameters, the runtime does increase substantially with the number of nodes, as well as with the number of auxiliary variables although to a lesser extent.
A closer inspection of the ``reference'' scenario 
suggests that the large grid values for the spike standard deviations $\nu_0$ consume the least computational time. Indeed, in such cases, a larger number of negligible coefficients are absorbed into the spike distribution, resulting in sparser model, which, in turn, promotes fast convergence \citep[see also,][]{wang2015scaling, rovckova2014emvs}. 
All the computations are implemented on an Intel Xeon CPU, 2.60 GHz.

\section{A study of gene regulation structure in monocyte}
\label{sec:application}
We return to the 
gene expression data presented in Section~\ref{sec:data}. Previous studies found substantial 
genetic regulation activity in monocytes on chromosome 12, likely linked with the development of immune-mediated diseases \citep{fairfax2012genetics, ruffieux2020global}. In particular, they identified a series of candidate ``hotspots'' (i.e., genetic variants controlling many gene levels) within $12$ megabase pairs of the lysozyme gene \emph{LYZ} (hereafter \emph{LYZ} region), which is known to have an important role in human innate immunity \citep{fairfax2012genetics, kim2014characterizing, ragland2017bacterial}. Here we undertake to clarify the mechanisms involved in the regulation activity of the \emph{LYZ} region by estimating the network dependence structure of genes controlled by the hotspots.

As motivated in Section~\ref{sec:data}, we aim to use our GMSS approach to estimate the partial correlation structures of genes in monocytes from the expression data described in \cite{fairfax2014innate}, taking advantage of gene-level auxiliary variables on the genetic regulation of these genes.  Specifically, to construct the auxiliary variables, we first analyse independent genetic and monocyte expression data from the CEDAR cohort \citep{momozawa2018ibd} with the Bayesian joint mapping approach \emph{atlasqtl} \citep{ruffieux2020global} and obtain posterior probabilities of association between each pair of genetic variant and gene. We next filter the genetic variants and genes, retaining the $29$ genetic variants, from the \emph{LYZ} region, associated with at least one gene, and the $137$ genes associated with at least one genetic variant from the \emph{LYZ} region, using a permutation-based Bayesian FDR of 20\% \citep[see][for details]{ruffieux2020global}. Finally, we define the auxiliary variables as the $P \times Q$ matrix $\bm{V}$ of pairwise posterior probabilities of association between the $P = 137$ genes and $Q = 29$ genetic variants. Note that this matrix is ``weakly sparse'' (most of its entries are close to zero), as most of the retained genetic variants are associated with one gene,
and very few are large hotspots, associated with about half of the genes.

We apply GMSS to two monocyte expression datasets from the \citet{fairfax2014innate} study, filtering the corresponding $P = 137$ genes. As introduced in Section~\ref{sec:data}, the first dataset involves genes quantified from unstimulated monocytes, for $N = 413$ subjects, and the second dataset involves genes quantified from monocytes that underwent stimulation by IFN-$\gamma$ inflammatory proxies, for $N = 366$ subjects. To evaluate the benefits of using the genetic association information in $\bm V$ with GMSS, we also compare our findings with those obtained from a classical spike-and-slab graphical model with no encoding of auxiliary variables, namely, using the GM$^*$ model. We do not consider GMN in this study for the statistical and computational reasons discussed in our simulations (Sections~\ref{sec:sim:accuracy}, \ref{sec:simulation:null} \& \ref{sec:sim:runtime}); in particular, (i) unlike GMSS, it doesn't entail a \emph{selection} of auxiliary variables, 
which could shed light on biological mechanisms, (ii) its edge-selection performance tends to be lower than that of GMSS, for a longer runtime, and (iii) it can lead to unstable inference due to insufficient regularisation.

The network sparsity estimated by the two approaches is similar, 3.6\% (GM$^*$) and 3.9\% (GMSS) for the unstimulated monocyte data, 4.1\% (GM$^*$) and 4.3\% (GMSS) for the stimulated monocyte data. The denser graphs in the latter case reflect previous findings showing that immune stimulation triggers substantial genetic activity, hence resulting in stronger gene dependence in the controlled networks.
In the unstimulated monocyte networks, a total of 312 edges are shared by the graphs inferred by the two approaches, 25 edges are unique to the  GM$^*$ graph, and 48 are unique to the GMSS graph. Similarly, in the stimulated monocyte networks, 346 edges are shared by the graphs inferred by the two approaches, 33 edges are unique to the  GM$^*$ graph, and 56 are unique to the GMSS graph.

\begin{figure}[ht!]
    \centering
    \includegraphics[width=\textwidth]{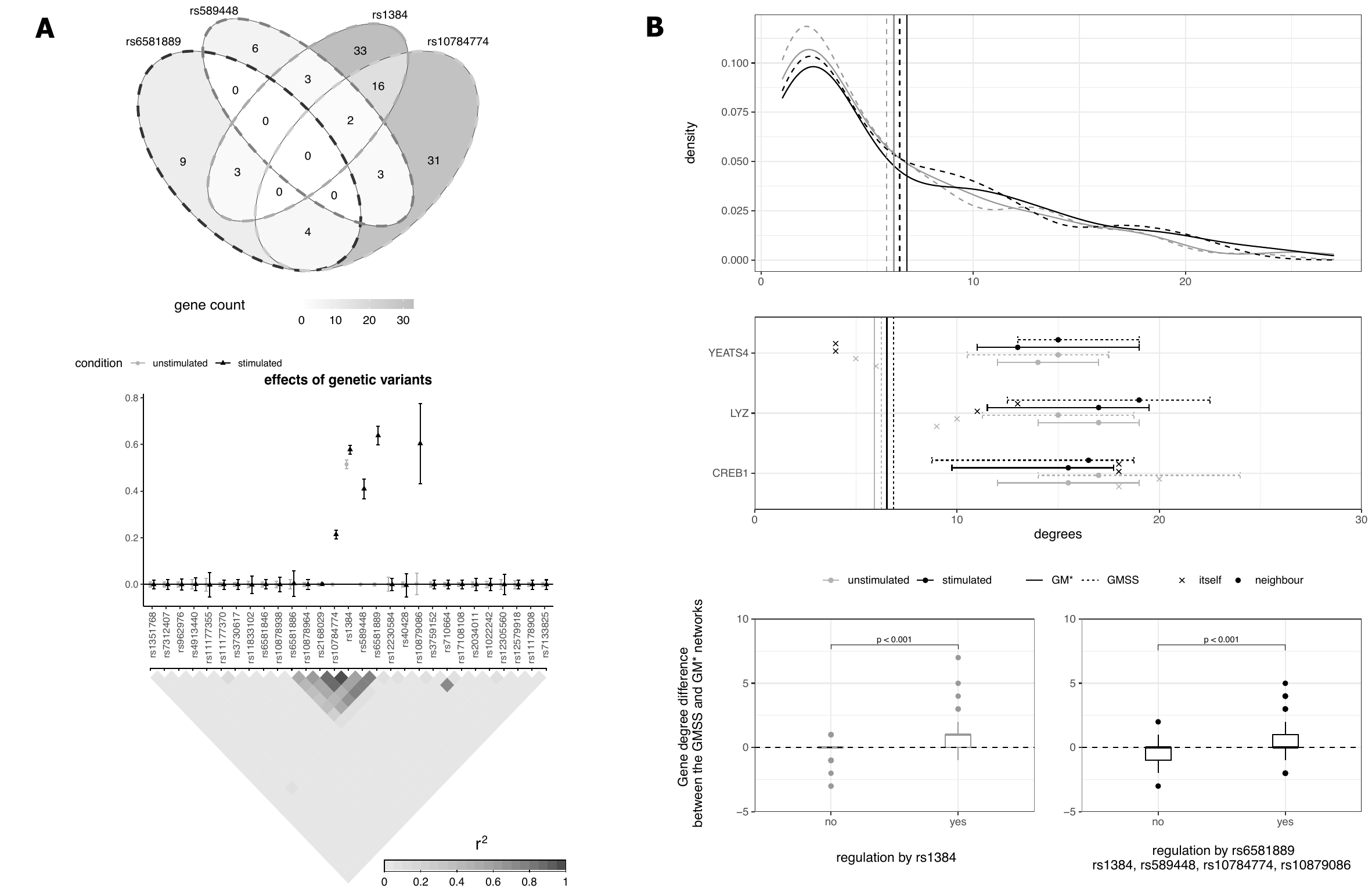}
    \caption{\footnotesize \normalfont \linespread{1.0}\selectfont Monocyte networks estimated by GM$^*$ and GMSS. (A) Top: number of genes associated with the four hotspot genetic variants picked by the GMSS in the stimulated monocyte network analysis in CEDAR dataset. Bottom: Estimated effects of auxiliary variables corresponding to genetic variants in the \emph{LYZ} region and their $r^2$ measure of linkage disequilibrium (genetic correlation), using GMSS in the unstimulated (grey) and simulated (black) monocyte network analyses along with their 95\% variational credible intervals. (B) Top: degree distributions of gene networks inferred by GM$^*$ (solid lines) and GMSS (dash lines), when monocytes are unstimulated (grey) and stimulated (black). Average degrees are indicated by vertical lines. 
    (B) Middle: degrees of \emph{LYZ}, \emph{YEATS4} and \emph{CREB1} using GM$^*$ (solid lines) and GMSS (dash lines), and median degree and interquartile range for their first-order neighbours. Bottom: difference in node degrees between the graphs inferred with GMSS and GM$^*$, for the gene sets categorised by the associations with genetic hotspot, rs1384, corresponding to the selected variable in the stimulated monocyte analysis, and genetic variants rs10879086, rs6581889, rs1384, rs589448, rs10784774, corresponding to the selected variables in the stimulated monocyte analysis. The $p$-values obtained from median-based permutation tests are shown. Genetic variants corresponding to the variables used as auxiliary data.}
    \label{fig:real_data}
\end{figure}

Inspecting the auxiliary variable PPIs estimated by GMSS indicates that the only variable retained is that corresponding to the genetic variant rs1384 
(using the median probability model rule PPIs $>$ 0.5), 
when monocytes are not stimulated. Interestingly, this variant was the second largest hotspot in the \emph{atlasqtl} analysis, where it was found to be associated with 57 genes at FDR 20\%; while a few of these genes are on the same chromosome as rs1384, most genes are spread across the entire genome. Moreover, one typically expects that any given block of highly correlated genetic variants (``locus'') will entail one or a few independent ``causal'' variants (if any), affecting disease traits \citep[see, e.g.,][]{schaid2018genome}. 
It is therefore not surprising that GMSS selects only one auxiliary variable, corresponding to a single hotspot genetic variant. In the stimulated monocyte analysis, however, GMSS selects five auxiliary variables. These include the variable corresponding to the hotspot rs1384, which is again found relevant to the graph structure. The remaining four variables correspond to genetic variants rs589448, rs6581889, rs10784774 and rs10879086, of which the first three are also hotspots, regulating $14$, $16$ and $56$ genes, respectively. Compared to the unstimulated setting, this larger number of selected variables may again reflect the greater genetic activity expected in response to the stimulation intervention. The estimated effects of the auxiliary variables are reported in Figure~\ref{fig:real_data}A (bottom). None of the variational credible intervals for the effects of the selected variables cover zero. Moreover, the variables corresponding to four hotspot genetic variants have smaller intervals, reflecting a higher level of certainty on their inclusion.
Finally, this figure also depicts the correlation structure, or \emph{linkage disequilibrium}, in the \emph{LYZ} region for the genetic variants corresponding to the auxiliary variables in $\bm V$. While rs10879086 displays very weak correlation with the other variants in the region, genetic variants rs10784774, rs1384, rs589448 and rs6581889 are in the same linkage disequilibrium block. However, they tend to control different sets of genes, as shown in Figure~\ref{fig:real_data}A (top).

To understand the effect of accounting for the genetic regulation information $\bm V$, we compare the degrees of the genes in the GMSS network with those in the GM$^*$ network which does not make use of this information. Figure~\ref{fig:real_data}B
indicates that the degree distributions inferred from GM$^*$ and GMSS are similar, regardless of the stimulation condition. 
However Figure~\ref{fig:real_data}B further shows that, in the unstimulated monocyte networks, the median degree of the set of genes regulated by the hotspot rs1384 
is higher in the GMSS estimation compared to the GM$^*$ estimation, and the increment is significant compared to random sets of unregulated genes of the same size using  permutation testing (empirical $p$-value~$<10^{-3}$). A similar observation holds in the stimulated monocyte networks, for the set of genes regulated by the five genetic variants corresponding to the auxiliary variables selected by GMSS. This suggests that the genetic regulation information is relevant to the gene dependence structure and that the genetic variants influencing this structure are effectively singled out and leveraged thanks to the GMSS top-level spike-and-slab submodel on the auxiliary variable effects.

The previous observations also suggest that inspecting genes with high degrees (``hubs'') may be particularly helpful to understand the mechanisms by which genes regulated by genetic hotspots influence immune-mediated diseases. The lists of genes with the top 10\% node degrees in the GMSS and GM$^*$ networks largely overlap, with 13
hub genes common to both methods in the unstimulated monocyte networks, and 14 in the stimulated monocyte networks (Supplementary Table~3).
The tripartite motif containing 16 like pseudogene (\emph{TRIM16L})
is the most central node in the unstimulated monocyte networks. Its degree is increased by 3 when auxiliary variables are exploited 
with the GMSS model (degree 27), compared to the GM$^*$ model (degree 24).
The tripartite motif (TRIM) is a large human protein family which has been shown to have a role in innate immune signalling pathways \citep{carthagena2009human, yang2020trim}. Although this motif still also appears among the top hubs in the stimulated monocyte networks, it has a lower degree (20 using both GM$^*$ and GMSS approaches), which aligns with the findings of
\cite{rajsbaum2008type}, reporting a down-regulation of TRIM16 in human macrophages stimulated with IFN-$\gamma$.
The top hub in the stimulated monocyte networks is the tumor suppressor p53-binding protein 2 (\emph{TP53BP2}). Its degree is increased by 4 in the GMSS network analysis (26) compared to the GM$^*$ analysis (22).
\cite{turnquist2014stat1} shows that IFN-$\gamma$ induces the expression of \emph{TP53BP2} (also called \emph{ASPP2}) in THP-1 cells, which are a type of human monocytic cell lines. \cite{xie2022dysregulated} found the deficiency of \emph{TP53BP2} causes dysregulation of both adaptive and innate immune systems, yet further research is required to fully understand its role in immune regulation.
These two examples suggest that GMSS likely helps pinpoint hub genes from important gene pathways, although their exact functions should be investigated further, for instance with experimental studies. Exploring the biological functions of other hubs may yield further valuable insights on the biological pathways implicated in diseases related to the innate immune system.

Finally, previous evidence suggests that hotspots from the \emph{LYZ} region regulate genes located nearby, which themselves mediate the expression of remote genes (possibly on other chromosomes). Specifically, the \emph{LYZ} gene itself and the YEATS domain containing 4 (\emph{YEATS4}) gene, both located within the \emph{LYZ} region, are candidate mediators thought to feed into the remote gene CAMP responsive element binding protein 1 (\emph{CREB1}), located on chromosome 2 \citep{fairfax2012genetics}. 
\cite{ruffieux2021epispot} further hypothesised that \emph{CREB1} 
may feed back onto \emph{LYZ}. Interestingly these genes, and their first-order neighbours, also have peculiar hub patterns in our analyses (Figure~\ref{fig:real_data}B). 
 Both \emph{LYZ} and \emph{CREB1} exhibit larger degrees than the mean network degree, in the unstimulated and stimulated monocyte estimations obtained with both GMSS and GM$^*$ (degrees ranging from $9$ to $20$). 
 In contrast, \emph{YEATS4} has a relatively small degree (ranging from $5$ to $7$ depending on the method and stimulation condition), 
 which may indicate that \emph{LYZ} and \emph{CREB1} are more promising candidate mediators than \emph{YEATS4}, aligning with the hypothesis of \cite{fairfax2012genetics}.
Furthermore, \emph{LYZ} and \emph{CREB1} are direct neighbours in both the GM$^*$ and GMSS networks (Supplementary Material Tables~4 and 5).
These observations lend additional support to  
the proposed hotspot-mediating role of \emph{LYZ} and \emph{CREB1}, and warrant further investigation about their mechanisms of action using experimental studies and independent data.

\section{Discussion}
\label{sec:discussion}
We have introduced a novel framework for estimating sparse dependence structures in Gaussian graphical models by leveraging external auxiliary information on the network structure. Our approach 
(i) uses a top-level spike-and-slab formulation to infer and encode the effects of node-level candidate variables;
(ii) implements a VBECM algorithm that outperforms existing ECM algorithms for graphical models while maintaining computational efficiency.
Our simulations and monocyte gene expression study suggest a substantial gain in accuracy in estimating graphical structures when auxiliary variables are exploited using our framework. 
Importantly, we have shown that our approach permits hypothesis-free selection of relevant variables, effectively discarding all variables irrelevant to the estimation of the underlying network. This permits supplying large panels of variables, with tens or possibly hundreds of candidate auxiliary variables, in an agnostic fashion, without risk of worsening inference. Moreover, the spike-and-slab PPIs are interpretable quantities for selecting edges, as well as auxiliary variables which may be relevant to the graph structure. Hence, our approach not only enhances statistical power to detect weak edges thanks to the auxiliary variables, but also pinpoints which variables are likely informative, thereby offering insights for interpreting the uncovered edges and formulating plausible mechanistic hypotheses in applied settings. 
Finally, we have shown that our VBECM algorithm, which infers full posterior distributions, improves upon the current deterministic ECM algorithms employed for fast network inference, without giving up on scalability. 

To conclude, we would like to emphasise the versatility of our framework, whose applicability and usefulness range across different problem settings and areas. First, as outlined in Section~\ref{sec:data}, other gene-level variables may aid in reconstructing gene expression networks. In particular, gene memberships to biological pathways can be encoded as binary node-level variables. This may not only improve the detection of gene structures – since genes in the same pathway tend to gather as modules in gene networks \citep{langfelder2007eigengene} – but also allow automatically identifying ``active'' pathways relevant to the network, which may point towards specific disease processes underlying the gene dependence structures. Clearly, our approach is not limited to the study of gene networks but is well suited to any molecular entities whose levels can be reasonably modelled using a multivariate Gaussian distribution; this includes, for instance, protein-protein interaction networks \citep{wang2016fastggm} or metabolite–metabolite association networks \citep{cakir2009metabolic}.
Our framework also has wide applications outside molecular research; to name only a few, it can be used to estimate (i) functional connectivity between brain regions \citep{belilovsky2016testing, li2018nonparametric}, for which features such as the size and brain lobe 
can serve as candidate auxiliary variables, (ii) relationships between diverse physical and mental health symptoms \citep{lafit2019partial}, encoding the mapping of the different symptoms with different diseases as auxiliary variables, (iii) historical social networks, such as collaboration networks \citep{newman2001structure, li2020high} or contact networks \citep{liljeros2001web, yang2021learning},
whose estimation may be improved by leveraging the characteristics of individuals, (iv) movement networks such as networks of road traffic states \citep{hara2018network}, along with auxiliary information on road conditions and properties, (v) financial networks representing stock prices within banking systems \citep{nicola2020information}, where factors like bank ownership structures and sizes may enhance the detection of network edges.

Different methodological extensions can be considered. First, a natural development would be to adapt the framework to joint inference across multiple related networks. In such a setting, one could envision having network-specific auxiliary variable effects, or even network-specific sets of auxiliary variables. For instance, the monocyte study could benefit from analysing genes collected from different cell types or tissues (hereafter ``conditions''), and the auxiliary variables could be constructed from genetic regulation information in each of the conditions concerned. Since it is expected that genetic regulation and gene relationships are partly shared across several conditions, estimation would benefit from exploiting common information across networks.
Such an extension could borrow ideas from existing proposals on Bayesian joint network inference, e.g., by \cite{peterson2015bayesian} \cite{li2019bayesian} or  \cite{lingjaerde2022scalable},
re-purposing them to the auxiliary variable setting. Furthermore, our modelling framework assumes a global influence of the selected auxiliary variables on the network structure. 
Modelling ``local effects'' on subsets of nodes may be more realistic in some applied settings.
A careful extension of the model hierarchy, possibly considering domain-specific knowledge, 
may prove useful to capture nuanced relationships within complex networks. 
In addition, our modelling framework can be easily adapted to incorporate edge-level, rather than node-level information 
through the probit submodel on the probability of edge inclusion.
Finally, it would be interesting to consider modelling variants based on alternative sparse prior distributions for edges and/or auxiliary variables, such as the horseshoe prior \citep{li2019graphical}, or the spike-and-slab lasso \citep{rovckova2018spike}
which is based on a mixture of double exponential distributions combines ideas of penalised regression and spike-and-slab shrinkage \citep{rovckova2018spike, bai2021spike}. 

Our framework serves as groundwork for more accurate, interpretable, versatile and scalable network estimation. 
We hope that it can pave the way to a more principled use of auxiliary variables in applied problems involving network inference, capitalising on the wealth of annotation sources that are now available to practitioners.

\subsection*{Software and reproducibility}
Our R package 
\texttt{navigm} (\underline{n}ode-level \underline{a}uxiliary \underline{v}ariables for \underline{i}mproved \underline{g}raphical \underline{m}odel inference) is available at \url{https://github.com/XiaoyueXI/navigm}. 
Replication code for all numerical examples presented in this article is available at \url{https://github.com/XiaoyueXI/navigm_addendum}.

\subsection*{Data}

The CD14$^+$ monocyte gene expression data presented in \citet{fairfax2014innate} have been generated using HumanHT-12v4 arrays and are freely available from ArrayExpress (accession E-MTAB-2232). 
The CEDAR dataset \citep{momozawa2018ibd} involves gene expression data from CD14$^+$ monocytes generated using Illumina HumanHT-12 v4 arrays and genotyping data generated using Illumina HumanOmniExpress-12 v1 A arrays. The raw expression data and genotype data are available from ArrayExpress (accession E-MTAB-6667 and E-MTAB-6666, respectively).

\subsection*{Acknowledgements}

We thank Sylvia Richardson for her valuable input during the development of the model, and Benjamin Fairfax for insightful discussions on the monocyte gene expression datasets. 
We also thank Colin Starr for his help in accessing computing resources. H.R. and X.X. would like to thank the Lopez–Loreta Foundation for their support.

\bibliographystyle{apalike}

\newpage

\begin{center}
{\large\bf SUPPLEMENTAL MATERIALS}
\end{center}
\beginsupplement
\addcontentsline{toc}{section}{Appendices}

\begin{appendices}
\section{Fast deterministic inference}\label{sec_app_det_inf}
\subsection{Algorithms}
In this section, we first adapt the expectation conditional maximisation (ECM) algorithm of \cite{li2019expectation} to our modelling framework and then present our variational Bayes expectation conditional maximisation (VBECM) algorithm for inference.

Recall the full model: for $N$ centred measurements, $\bm{Y} = (\bm{y_1}, \ldots, \bm{y_N})$, and $Q$ auxiliary variables, $\bm{V} = \left(V_1^T, \ldots, V_Q^T\right)^T$ for $P$ nodes in a graph,
\begin{eqnarray} \nonumber
\bm{y_n} & \overset{\text{iid}}{\sim}& \mathcal{N}_P(\bm 0, \bm \Omega^{-1}), \quad \bm{\Omega} \in \mathcal{M}^+, \quad n = 1, \ldots, N,\\ \nonumber
\omega_{ii}  &\sim& \mathrm{Exp}\left({\lambda/2}\right), \quad i = 1, \ldots, P,\\ \nonumber
\omega_{ij}\mid \delta_{ij}, \tau &\sim& \delta_{ij}\mathcal{N}\left(0, \nu_1^2/\tau \right) + (1-\delta_{ij}) \mathcal{N}\left(0, \nu_0^2/\tau\right),\quad  \nu_0 \ll \nu_1, \quad 1 \leq i<j \leq P, \\ \label{eq:app:probit}
\delta_{ij} \mid \rho_{ij} &\sim& \text{Bern}(\rho_{ij}), \quad 
\rho_{ij} = \Phi(\alpha_{ij}), \quad \alpha_{ij} = \zeta + \bm{v_i \beta} + \bm{v_j \beta}, \\ \nonumber
\beta_q \mid \gamma_q, \sigma^2 &\sim&  \gamma_{q}\mathcal{N}(0,\sigma^2) + (1-\gamma_{q}) \delta(\beta_q), \quad
\gamma_q\mid o \sim \text{Bern}(o), \quad q=1,\ldots, Q,\\ \nonumber
\zeta &\sim& \mathcal{N}(n_0, t_0^2), \quad o \sim  \text{Beta}(a_o, b_o), \quad
\sigma^{-2} \sim \text{Gamma}(a_\sigma, b_\sigma),
\end{eqnarray}
where $\mathcal{M}^+$ is the set of symmetric positive definite matrices, $\Phi(\cdot)$ is the standard normal cumulative distribution function and $\delta(\cdot)$ is the Dirac distribution.
To obtain analytical updates, we reparameterise the probit link formulation \eqref{eq:app:probit}
using the classical data-augmentation which introduces a latent vector $\bm{z}$ as follows
\begin{eqnarray*}
\delta_{ij} \mid z_{ij}&=& \mathds{1}\{ z_{ij} >0 \}, \quad z_{ij} \mid \zeta, \bm{\beta}\sim \mathcal{N} \left( \alpha_{ij}, 1 \right), \quad \alpha_{ij} = \zeta + \bm{v_i \beta} + \bm{v_j \beta}.
\end{eqnarray*}

\subsubsection{Derivation of the expectation conditional maximisation algorithm}\label{app:emdev}
The ECM algorithm, initially proposed by
\cite{meng1993maximum}, introduces several simpler conditional maximisation steps to replace a complicated maximisation step for which no analytical form can be obtained. \cite{li2019expectation} were the first to use the ECM algorithm for the spike-and-slab graphical model. They employed a continuous spike-and-slab prior to ensure closed-form updates, which is achieved by repurposing the Gibbs sampling slicing procedure proposed by \cite{wang2015scaling}. 

To obtain an ECM implementation of our approach, we replace the discrete spike-and-slab for the node-level auxiliary variable selection in~\eqref{eq:codata_selection} of the main text,
\begin{eqnarray*}
\beta_q \mid \gamma_q, \sigma^2 &\sim&  \gamma_{q}\mathcal{N}(0,\sigma^2) + (1-\gamma_{q}) \delta(\beta_q), \quad \sigma^{-2} \sim \text{Gamma}(a_\sigma, b_\sigma),
\end{eqnarray*}
with a continuous formulation, as in \eqref{eq:cssprior},
\begin{eqnarray}
\label{eq:app:css_var}
 \beta_q \mid \gamma_q, \tau_2 &\sim&  \gamma_{q}\mathcal{N}\left(0,\sigma_1^2/\tau_2 \right) + (1-\gamma_{q}) \mathcal{N}\left(0,\sigma_0^2/\tau_2 \right), \quad
 \tau_2 \sim \text{Gamma}(a_\sigma, b_\sigma),
\end{eqnarray}
for $q=1,\ldots, Q$, where $\sigma_0, \sigma_1 > 0$ are set to small and large values respectively (see Section~\ref{sec:practical:ssvar} in the main text) and and $\tau_2$ is a scaling parameter. In what follows we also denote by $\tau_1$, the scale parameter $\tau$ in the bottom-level continuous spike-and-slab formulation for the edge selection in~\eqref{eq:cssprior} of the main text.

We introduce the notation $\bm{\Theta} = (\bm{\Theta_1}, \bm{\Theta_2})$, with $\bm{\Theta_1} = (\tau_1, \tau_2, \zeta, \bm{\beta},o)$ representing unknown parameters and $\bm{\Theta_2} = (\bm{\delta}, \bm{z},\bm{\gamma})$ representing latent variables. 
Given the estimates from the previous iteration $t$, the objective function to optimise is 
\begin{eqnarray*}
Q(\bm{\Omega},\bm{\Theta_1} \mid \bm{\Omega}^{(t)},\bm{\Theta_1}^{(t)}) & = &\mathbb{E}_{\bm{\Theta_2} \mid \bm{\Omega}^{(t)},\bm{\Theta_1}^{(t)}, \bm{Y}} \left\{ \log p (\bm{\Omega},\bm{\Theta_1},\bm{\Theta_2} \mid \bm{Y}) \mid \bm{\Omega}^{(t)},\bm{\Theta_1}^{(t)}, \bm{Y} \right\} \\
& = & \frac{N}{2} \log(|\bm{\Omega}|) - \frac{1}{2} \text{tr}(\bm{Y^T Y} \bm{\Omega}) -  \frac{\lambda}{2}\sum_{i=1}^P\omega_{ii} \\
& & - \log(\nu_1) \sum_{i<j} \mathbb{E}_{\cdot | \cdot}(\delta_{ij}) - \log(\nu_0) \sum_{i<j} \left\{1-\mathbb{E}_{\cdot | \cdot}(\delta_{ij}) \right\} + \frac{P(P-1)}{4} \log (\tau_1)\\
& & - \frac{\tau_1}{2} \sum_{i<j} \omega_{ij}^2 \mathbb{E}_{\cdot|\cdot}\left(\frac{\delta_{ij}}{\nu_1^2}  +  \frac{1- \delta_{ij}}{\nu_0^2} \right)  + (a_\tau-1)\log(\tau_1) - b_\tau \tau_1\\
& & + \sum_{i<j} \mathbb{E}_{\cdot | \cdot} \bigg[ \delta_{ij} \log \mathds{1}\{z_{ij}>0\} + (1-\delta_{ij})\log \mathds{1}\{z_{ij}\leq 0\} \bigg]\\ \nonumber
&  & - \frac{1}{2} \sum_{i<j} \bigg\{ \mathbb{E}(z_{ij}^2) - 2\alpha_{ij} \mathbb{E}(z_{ij})  + \alpha_{ij}^2 \bigg\}  -\frac{\zeta^2}{2 t_0^2} + \frac{n_0 \zeta}{t_0^2}\\
& & -  \log(\sigma_1) \sum_{q=1}^Q \mathbb{E}_{\cdot|\cdot} (\gamma_q) - \log(\sigma_0) \sum_{q=1}^Q \left\{1 - \mathbb{E}_{\cdot|\cdot} (\gamma_q) \right\} + \frac{Q}{2} \log(\tau_2)\\
&  & -\frac{\tau_2}{2} \sum_{q=1}^Q \beta_q^2 \mathbb{E}_{\cdot\mid\cdot} \left( \frac{\gamma_q}{\sigma_1^2}  +  \frac{1-\gamma_q}{\sigma_0^2} \right) + (a_\sigma-1)\log(\tau_2)  - b_\sigma \tau_2 \\
& & + \log(o) \sum_{q=1}^Q \mathbb{E}_{\cdot\mid \cdot}(\gamma_q)  +  \log(1-o) \sum_{q=1}^Q \left\{1- \mathbb{E}_{\cdot\mid \cdot}(\gamma_q)  \right\}  \\ \nonumber
&  & + (a_o-1) \log o + (b_o-1) \log(1-o)  + \text{cst}, \nonumber
\end{eqnarray*}
\normalsize
where $\mathbb{E}_{\cdot\mid\cdot} (\cdot):= \mathbb{E}_{\bm{\Theta_2} \mid \bm{\Omega}^{(t)}, \bm{\Theta_1}^{(t)}, \bm{Y}} (\cdot)$ denotes the expectation of the posterior conditional on current estimates and observations and \text{cst} is constant with respect to $\bm{\Omega}$ and $\bm{\Theta_1}$. 
The ECM algorithm iterates between an ``expectation step'' (E-step), which evaluates the conditional expectations in the objective function, and a ``conditional maximisation step'' (CM-step), which solves for mode of the objective function in a coordinate ascent manner, until the objective function converges. These steps are detailed below. 

\paragraph*{The E-step}
The conditional expectations in the objective function are
\begin{eqnarray*}
\mathbb{E}_{\cdot\mid\cdot} ( \delta_{ij} ) &=& \frac{p\left(\omega_{ij}^{(t)}\mid\delta_{ij}=1\right)\Phi\left
(\alpha_{ij}^{(t)}\right) }{p\left(\omega_{ij}^{(t)}\mid\delta_{ij}=1\right)\Phi\left(\alpha_{ij}^{(t)}\right)  + p\left(\omega_{ij}^{(t)}\mid\delta_{ij}=0\right) \bigg\{1 - \Phi\left(\alpha_{ij}^{(t)}\right)\bigg\} }, \\
\mathbb{E}_{\cdot\mid\cdot} ( \gamma_q ) &=& \frac{p\left(\beta_q^{(t)}\mid\gamma_q=1\right) o^{(t)} }{p\left(\beta_q^{(t)}\mid\gamma_q=1\right) o^{(t)}  + p\left(\beta_q^{(t)}\mid\gamma_q=0\right) \left( 1- o^{(t)} \right) },
\end{eqnarray*}
and 
\begin{eqnarray*}
\mathbb{E}_{\cdot\mid\cdot} \left( \frac{\delta_{ij}}{\nu_1^2}  +  \frac{1- \delta_{ij}}{\nu_0^2}  \right) = \frac{\mathbb{E}_{\cdot\mid\cdot} ( \delta_{ij}) }{\nu_1^2} +\frac{1- \mathbb{E}_{\cdot\mid\cdot} ( \delta_{ij} )}{\nu_0^2}:= d_{ij}^* , \\
\mathbb{E}_{\cdot\mid\cdot} \left( \frac{\gamma_q}{\sigma_1^2}  +  \frac{1-\gamma_q}{\sigma_0^2}  \right) =  \frac{\mathbb{E}_{\cdot\mid\cdot} ( \gamma_q ) }{\sigma_1^2} +\frac{1- \mathbb{E}_{\cdot\mid\cdot} ( \gamma_q ) }{\sigma_0^2} := g_q^*.
\end{eqnarray*}
Moreover, we have
\begin{eqnarray*}
\mathbb{E}_{\cdot\mid\cdot}(z_{ij}) & = & \alpha_{ij}^{(t)}+M\left(\alpha_{ij}^{(t)},1\right)\mathbb{E}_{\cdot\mid\cdot} ( \delta_{ij} ) +M\left(\alpha_{ij}^{(t)},0\right)\Big\{1-\mathbb{E}_{\cdot\mid\cdot} ( \delta_{ij} ) \Big\} \\
& = & \alpha_{ij}^{(t)} + \mathbb{E}_{\cdot\mid\cdot} ( \delta_{ij} ) \bigg\{ M\left(\alpha_{ij}^{(t)},1\right)-M\left(\alpha_{ij}^{(t)},0\right)\bigg\} + M\left(\alpha_{ij}^{(t)},0  \right), \\
\mathbb{E}_{\cdot \mid \cdot}(z_{ij}^2)
&=& \mathbb{E}_{\cdot\mid\cdot} ( \delta_{ij} )\bigg\{ \left(\alpha_{ij}^{(t)}\right)^2 + 1 + \alpha_{ij}^{(t)} M\left(\alpha_{ij}^{(t)},1\right)\bigg\} \\
& & +\Big\{1-\mathbb{E}_{\cdot\mid\cdot} ( \delta_{ij} ) \Big\} \bigg\{\left(\alpha_{ij}^{(t)} \right)^2 + 1 + \alpha_{ij}^{(t)}M\left(\alpha_{ij}^{(t)},0\right) \bigg\}\\
& = &
\alpha_{ij}^{(t)} \mathbb{E}_{\cdot\mid\cdot}(z_{ij})  + 1,
\end{eqnarray*}
where $M(\mu,\gamma)$ is the inverse Mills ratio,
\begin{equation*}
M(\mu,\gamma) = (-1)^{1-\gamma} \frac{\varphi(\mu)}{{\Phi(\mu)}^\gamma \big\{
1-\Phi(\mu)\big\}^{1-\gamma}}, \quad \mu \in \mathbb{R}, \quad r = 0, 1,
\end{equation*}
$\varphi(\cdot)$ is the standard normal density and $\Phi(\cdot)$ is the standard normal cumulative distribution function. 
Last, since $\delta_{ij} = 1$ when $z_{ij}>0$ and 0 otherwise, we always have 
\begin{eqnarray*}
\delta_{ij} \log \mathds{1}\{z_{ij}>0\} + (1-\delta_{ij})\log \mathds{1}\{z_{ij}\leq 0\} = 0,
\end{eqnarray*}
and its conditional expectation is thus zero.

\paragraph*{The CM-step}
The CM-step finds the posterior mode by maximising the objective function with respect to the unknown parameters $\bm{\Omega}$ and $\bm{\Theta_1}$ in a coordinate ascent manner. The closed-form updates for the scale parameters $\tau_1$ and $\tau_2$ are
\begin{eqnarray*} \label{eq:tau1}
\tau_1^{(t+1)} = \frac{ P(P-1)/2 + 2a_\tau - 2}{\sum_{i<j} \omega_{ij}^2 d_{ij}^* + 2b_\tau}, \quad 
\tau_2^{(t+1)} = \frac{Q + 2a_\sigma -2 }{\sum_{q=1}^Q \beta_q ^2 g_q^* + 2b_\sigma}.
\end{eqnarray*}
The update for $\zeta$ is 
\begin{eqnarray*}
\zeta^{(t+1)} = \frac{ 2 n_0 + 2t_0^2 \sum_{i<j} \mathbb{E}_{\cdot\mid\cdot}(z_{ij}) - 2t_0^2 \sum_{i<j} \sum_q \beta_q (V_{iq} + V_{jq})}{P(P-1)t_0^2+ 2}.
\end{eqnarray*}
The update for $\beta_q$ is given by
\begin{eqnarray*} 
\beta_q^{(t+1)} & = & \frac{\sum_{i<j} (V_{iq} + V_{jq})\mathbb{E}_{\cdot\mid\cdot}(z_{ij}) -  \zeta (P-1) \sum_{i=1}^P V_{iq}}{(P-1) \sum_{i=1}^P V_{iq}^2 + 2 \sum_{i<j} V_{iq} V_{jq} + \tau_2 g_q^*} \\
&  & - \frac{(P-1)\sum_{i=1}^P V_{iq} \sum_{q^\prime\neq q} V_{iq^\prime} \beta_{q^\prime}- \sum_{i<j} \sum_{q^\prime \neq q} \beta_{q^\prime} (V_{iq} V_{jq^\prime} +V_{iq^\prime} V_{jq})}{{(P-1) \sum_{i=1}^P V_{iq}^2 + 2 \sum_{i<j} V_{iq} V_{jq} + \tau_2 g_q^*}},
\end{eqnarray*}
for $q=1,\ldots, Q$.
The update for $o$ is 
\begin{eqnarray*}
o^{(t+1)} &=& \frac{\sum_{q=1}^Q \mathbb{E}_{\cdot\mid\cdot} ( \gamma_q ) + a_o -1}{Q + a_o + b_o  -2}.
\end{eqnarray*}
There is no closed form for the joint update of the precision matrix $\bm{\Omega}$. However,  \cite{wang2015scaling} provides analytical block updates, namely, by updating each column (and thus each row due to symmetry) of $\bm \Omega$ and iterating over all the columns, 
\begin{eqnarray*}
\bm{\omega}_{-ii} &=& \left(\bm{\omega}_{i-i}\right)^T = - \bigg\{ (s_{ii} + \lambda) \left( \bm{\Omega}_{-i-i} \right)^{-1}+ \text{diag}\left(\tau_1 \bm{d}_{-ii}^* \right) \bigg\}^{-1} \bm{S}_{-ii}, \\ \nonumber
\omega_{ii} &=& \bm{\omega}_{i-i} \left(\bm \Omega_{-i-i}\right)^{-1} \bm{\omega}_{-ii}+ \frac{N}{s_{ii}+ \lambda},
\end{eqnarray*}
where $\bm{S} = \bm{Y^T Y}$, $\bm{d}^*$ is a $P\times P$ matrix with entries $d_{ij}^*$,  and the subscript $-i$ refers to the removal of $i$th row (the first subscript) or column (the second subscript).
\subsubsection{Derivation of the variational expectation conditional maximisation algorithm}\label{app:vbemdev}
Next, we detail the VBECM algorithm for GMSS, which allows us to approximate the full parameter posterior distributions instead of targeting the posterior modes. 
We consider the mean-field approximation,
\begin{eqnarray*}
q(\bm{\Omega}, \bm{\Theta}) &=& q(\bm{\Omega}) \prod_{i<j}q(\delta_{ij}, z_{ij}) q(\tau) q(\zeta) \prod_q q(\beta_q, \gamma_q) q(o) q(\sigma^2),
\end{eqnarray*}
where $\bm{\Theta} = (\bm{\delta},\bm{z}, \tau,\zeta, \bm{\beta}, \bm{\gamma}, \sigma^2, o)$. Since the auxiliary variables $z_{ij}$ fully determine the binary variables $\delta_{ij}$, we group $\delta_{ij}$ and $z_{ij}$ in one mean field factor. 
Then we find the optimal variational posterior by maximising the evidence lower bound (ELBO; \eqref{eq:elbo} of the main text) in a coordinate ascent fashion; the optimal solution for the $k$th factor in the mean-field approximation, denoted by $\bm\theta_k$, is given by 
\begin{eqnarray*}
q(\bm\theta_k) \propto \exp\bigg\{ \mathbb{E}_{q(\bm\theta_{-k})} \log p(\bm{Y}, \bm{\Omega}, \bm{\Theta} ) \bigg\},
\end{eqnarray*}
where $\bm\theta_{-k} = \bm{\Theta} \setminus \bm\theta_k$ \citep{bishop2006pattern}. 
The variational distribution is known in closed form except for $\bm{\Omega}$,
but the conditional posterior mode of each column is known and thus can be updated by a CM-step. 

The VBECM algorithm alternates between a ``variational expectation step'' (VBE-step), which approximates the posterior distribution (and thus the expectation) of all the parameters except $\bm \Omega$, and a CM-step, which evaluates the conditional maximisation for $\bm \Omega$, until the ELBO converges. To approximate the posterior distribution of parameters in the VBE-step, we iterate over each partition of parameters and optimise the ELBO until convergence.
\paragraph*{The VBE-step}
The optimal variational distribution of $\delta_{ij}$ and $z_{ij}$ satisfies 
\begin{eqnarray*} \label{eq:z}
q(\delta_{ij}, z_{ij}) & = &q(z_{ij} \mid \delta_{ij}) q(\delta_{ij}), \quad 1 \leq i < j \leq P,
\end{eqnarray*}
with
\begin{eqnarray*} 
z_{ij} \mid \delta_{ij}, \bm{Y} &\sim&
\mathcal{TN}\bigg\{\alpha_{ij}^{(1)}, 1, (-1)^{1-\delta_{ij}} z_{ij} > 0\bigg\}, \quad \delta_{ij}\mid \bm{Y} \sim \text{Bern}\left(\delta_{ij}^{(1)}\right),
\end{eqnarray*}
where $\mathcal{TN}$ denotes a truncated normal distribution, the superscripts $^{(1)}$ and $^{(2)}$ denote the first, respectively second moments of the variable they are applied to, 
\begin{eqnarray*}
\alpha_{ij}^{(1)}&  = &\zeta^{(1)} + \sum_q V_{iq}\beta_q^{(1)} + \sum_q V_{jq}\beta_q^{(1)}, \\
\frac{1}{\delta_{ij}^{(1)} }&=& 1+ \exp \Bigg[ \log\left( \frac{\nu_1}{\nu_0}\right) + \frac{\tau^{(1)} \omega_{ij}^{2}}{2} \left( \frac{1}{\nu_1^2} - \frac{1}{\nu_0^2}\right)   + \log\bigg\{ 1-\Phi\left(\alpha_{ij}^{(1)} \right) \bigg\} - \log  \Phi \left(\alpha_{ij}^{(1)} \right)  \Bigg].
\end{eqnarray*}
The first two moments of
$z_{ij}$ are 
\begin{eqnarray*}
z_{ij}^{(1)} & = & \alpha_{ij}^{(1)} +  M\left(\alpha_{ij}^{(1)}, 0\right) + \delta_{ij}^{(1)} \bigg\{  M\left(\alpha_{ij}^{(1)}, 1\right) -  M\left(\alpha_{ij}^{(1)}, 0\right)\bigg\}, \quad  z_{ij}^{(2)} = \alpha_{ij}^{(1)} z_{ij}^{(1)}+ 1.
\end{eqnarray*}
Then, we find 
\begin{eqnarray*}\label{eq:tau}
\tau \mid \bm{Y} &\sim& \text{Gamma}(\alpha_\tau, \beta_\tau),
\end{eqnarray*}
with
\begin{eqnarray*}
\alpha_\tau &=&  \frac{P(P-1)}{4}+ a_\tau, \quad \beta_\tau = \frac{1}{2} \sum_{i<j} \omega_{ij}^2 \left( \frac{\delta_{ij}^{(1)}}{\nu_1^2} + \frac{1-\delta_{ij}^{(1)}}{\nu_0^2}\right) + b_\tau , \\
\tau^{(1)} &=& \alpha_\tau/\beta_\tau, \quad (\log \tau)^{(1)} = \psi(\alpha_\tau) - \log(\beta_\tau),
\end{eqnarray*}
where $\psi(x)$ stands for the digamma function which is defined by the gamma function and its derivative, $\psi(x) = \Gamma^{\prime}(x)/\Gamma(x)$ for $x > 0$. Similarly, the optimal variational distribution of $\beta_q$ and $\gamma_q$~is
\begin{eqnarray*}
q(\beta_q, \gamma_q) & = & q(\beta_q\mid \gamma_q)q(\gamma_q), \quad q = 1,\ldots, Q,
\end{eqnarray*}
with
\begin{eqnarray*}
\beta_q \mid \gamma_q, \bm{Y} = 1 &\sim& \mathcal{N}(\mu_{\beta,q}, \sigma_{\beta,q}^2), \quad
\beta_q \mid \gamma_q, \bm{Y} = 0 \sim \delta(\beta_q), \quad \gamma_q \mid \bm{Y} \sim \text{Bern}\left(\gamma_q^{(1)}\right),
\end{eqnarray*}
where
\begin{eqnarray*}
\mu_{\beta,q} 
& = & \sigma_{\beta,q}^{2} \bigg\{ \sum_{i<j} (V_{iq} + V_{jq}) z_{ij}^{(1)} - 
(P-1) \zeta^{(1)} \sum_{i=1}^P V_{iq}  \\
& &  - (P-1)  \sum_{i=1}^P V_{iq} \sum_{q^\prime \neq q} V_{iq^\prime} \beta_{q^{\prime}}^{(1)} - \sum_{i<j}\sum_{q^\prime \neq q} (V_{iq} V_{jq^{\prime}} + V_{iq^\prime} V_{jq})\beta_{q^{\prime}}^{(1)} \bigg\},\\
\sigma^{-2}_{\beta,q} & = &  \left( \sigma^{-2} \right)^{(1)} + (P-1)\sum_{i=1}^P V_{iq}^2 + 2\sum_{i<j} V_{iq} V_{jq}, 
\end{eqnarray*}
and 
\begin{eqnarray*}
\frac{1}{\gamma_q^{(1)}} & = & 1 + \exp \bigg[  \Big\{ \log(1-o) \Big\}^{(1)} - \Big\{ \log(o) \Big\}^{(1)} - \frac{1}{2} \left(\log \sigma^{-2}\right)^{(1)} - \frac{\mu_{\beta,q}^2}{2\sigma_{\beta,q}^2} + \frac{1}{2} \log \sigma_{\beta,q}^{-2} \bigg].
\end{eqnarray*}
The first two moments of $\beta_q$ are 
\begin{eqnarray*}
\beta_q^{(1)} = \gamma_q^{(1)} \mu_{\beta, q}, \quad \beta_q^{(2)} = \gamma_q^{(1)} \left(\mu^2_{\beta, q} + \sigma^2_{\beta, q}\right).
\end{eqnarray*}
Also, we have
\begin{eqnarray*}
\zeta \mid \bm{Y} \sim \mathcal{N}(\mu_\zeta, \sigma_\zeta^2), 
\end{eqnarray*}
where 
\begin{eqnarray*}
\mu_\zeta & = & \sigma_\zeta^2 \Bigg\{\sum_{i<j} z_{ij}^{(1)} - \sum_q\beta_q^{(1)} \sum_{i<j} (V_{iq} + V_{jq}) + \frac{n_0}{t_0^2} \Bigg\}, \\
\sigma_\zeta^{-2} &=& \frac{1}{t_0^2} + \frac{P(P-1)}{2} , \quad
\zeta^{(1)}  =  \mu_\zeta, \quad \zeta^{(2)} = \mu_\zeta^2 + \sigma_\zeta^2.
\end{eqnarray*}
 The optimal variational distribution of $\sigma^{-2}$ is 
\begin{eqnarray*}\label{eq:sigma}
\sigma^{-2} \mid \bm{Y} &\sim& \text{Gamma}(\alpha_\sigma, \beta_\sigma), 
\end{eqnarray*}
where
\begin{eqnarray*}
\alpha_\sigma & = & \frac{1}{2}\sum_q \gamma_q^{(1)} + a_\sigma, \quad
\beta_\sigma  =  \frac{1}{2}\sum_q \gamma_q^{(1)} \beta_q^{(2)} + b_\sigma , \\
\left( \sigma^{-2} \right)^{(1)} & = &  \alpha_\sigma/\beta_\sigma, \quad
\Big\{ \log \left(\sigma^{-2}\right) \Big\}^{(1)} = \psi(\alpha_\sigma) - \log(\beta_\sigma).
\end{eqnarray*}
Finally, we find
\begin{eqnarray*}
o \mid \bm{Y} \sim \text{Beta}(\alpha_{o}, \beta_{o}),
\end{eqnarray*}
where
\begin{eqnarray*}
\alpha_{o} & = & \sum_q \gamma_q^{(1)} + a_o, \quad
\beta_{o}  =  \sum_q\left(1 - \gamma_q^{(1)}\right) + b_o, \quad o^{(1)}  =  a_{o}/(a_{o}+ b_{o}), \\
\Big\{ \log(o) \Big\}^{(1)} &=& \psi(\alpha_{o}) - \psi(\alpha_{o} + \beta_{o}) , \quad
\Big\{ \log(1-o) \Big\}^{(1)} = \psi(\beta_{o}) - \psi(\alpha_{o} + \beta_{o}).
\end{eqnarray*}
\paragraph*{The CM-step}
The variational distribution of $\bm{\Omega}$ is not known in closed form and entails the constraint of symmetry and positive definiteness. However, the conditional distribution of its last column is tractable \citep{wang2015scaling}, allowing for conditional maximisation of the last column while holding the rest of the matrix entries fixed. We therefore can reframe this by considering the variational distribution of $\bm \Omega$ to a Dirac delta distribution, which is zero except for an unknown value to be estimated. This approach is equivalent to the CM-step in the ECM algorithm. For further details, we refer to the ECM derivation.

\paragraph*{The ELBO}
The ELBO is given by 
\begin{eqnarray*}
\mathcal{L}(q) & = &\mathbb{E}_{q(\bm{\Omega, \Theta})} \Big\{ \log p(\bm{Y},\bm{\Omega}, \bm{\Theta}) \Big\} - \mathbb{E}_{q(\bm{\Omega, \Theta})} \Big\{ \log  q(\bm{\Omega}, \bm{\Theta}) \Big\} \\
&=& \mathcal{L}_{\bm{Y}}(\bm{Y} \mid \bm{\Omega}) + \mathcal{L}_{\bm\Omega}(\bm{\Omega}\mid \bm{\delta},\tau) + \sum_{i<j} \mathcal{L}_{\bm{\delta, z}} (\delta_{ij}, z_{ij} \mid \zeta, \bm{\beta}) + \mathcal{L}_{\tau} (\tau) \\ \nonumber
&& + \mathcal{L}_{\zeta} (\zeta) + \sum_q    \mathcal{L}_{\bm{\beta}, \bm{\gamma}} (\beta_q, \gamma_q \mid o, \sigma^2) + \mathcal{L}_{o} (o) + \mathcal{L}_{\sigma} (\sigma^2), 
\end{eqnarray*}
where 
\begin{eqnarray*}
\mathcal{L}_{\bm{Y}}(\bm{Y}\mid \bm{\Omega}) & =& \log p(\bm{Y}\mid \bm{\Omega})= \frac{N}{2} \log|\bm{\Omega}| - \frac{1}{2} \text{tr}(\bm{Y}^T \bm{Y} \bm{\Omega}),\\
\mathcal{L}_{\bm\Omega}(\bm{\Omega}\mid \bm{\delta},\tau) & = & \mathbb{E}_{q(\bm{\Omega, \Theta})}  \bigg\{\sum_{i=1}^P\log p(\omega_{ii}) + \sum_{i<j} \log p(\omega_{ij} \mid  \delta_{ij}, \tau)\bigg\} \\
& = & -  \frac{\lambda}{2}\sum_{i=1}^P\omega_{ii} - \log \nu_1 \sum_{i<j} \delta_{ij}^{(1)} - \log \nu_0 \sum_{i<j} \left(1-\delta_{ij}^{(1)}\right)  \\
& & - \frac{\tau^{(1)}}{2}\sum_{i<j}  \omega_{ij}^{2} \left( \frac{\delta_{ij}^{(1)}}{\nu_1^2} + \frac{1-\delta_{ij}^{(1)}}{\nu_0^2}\right)  + \frac{P(P-1)}{4} \left( \log \tau \right)^{(1)}, \\
\end{eqnarray*}
\begin{eqnarray*}
\mathcal{L}_{\bm{\delta},\bm{z}} (\delta_{ij}, z_{ij} \mid \zeta, \bm{\beta}) &=& \mathbb{E}_{q(\bm{\Omega, \Theta})} \Big\{ \log p(\delta_{ij} \mid z_{ij}) + \log p(z_{ij}\mid \zeta, \bm{\beta}) - \log q(z_{ij} \mid \delta_{ij}) - \log q(\delta_{ij}) \Big\}\\
&=&  \Big[ \delta_{ij} \log \mathds{1}\{z_{ij}\geq 0\}\Big]^{(1)} + \Big[ \left(1-\delta_{ij}\right) \log \mathds{1}\{z_{ij}< 0\}\Big]^{(1)} \\
& & - \frac{1}{2} \left( z_{ij}^{(2)} - 2\alpha_{ij}^{(1)} z_{ij}^{(1)} + \alpha_{ij}^{(2)} \right) - \mathbb{E}_{q(\bm{\Omega, \Theta})}\log q(z_{ij}\mid \delta_{ij}) -\mathbb{E}_{q(\bm{\Omega, \Theta})} \log q(\delta_{ij}),
\end{eqnarray*}
with 
\begin{eqnarray*}
\mathbb{E}_{q(\bm{\Omega, \Theta})}\log q(z_{ij}\mid \delta_{ij}) & = & - \frac{1}{2} \bigg\{ z_{ij}^{(2)} - 2 z_{ij}^{(1)}\alpha_{ij}^{(1)}+ \left( \alpha_{ij}^{(1)} \right)^2 \bigg\}  \\
& &
- \delta_{ij}^{(1)} \log\bigg\{ \Phi \left(\alpha_{ij}^{(1)}\right) \bigg\} - \left( 1 - \delta_{ij}^{(1)}\right) \log\bigg\{ 1 - \Phi\left(\alpha_{ij}^{(1)}\right) \bigg\} \\
& & + \delta_{ij}^{(1)} \Big[ \log \mathds{1}\{z_{ij} >0 \} \Big]^{(1)} + \left(1-\delta_{ij}^{(1)}\right)  \Big[ \log \mathds{1}\{z_{ij} \leq 0\}  \Big]^{(1)},
\end{eqnarray*}
using the definition of conditional entropy and 
\begin{eqnarray*}
\mathbb{E}_{q(\bm{\Omega, \Theta})}\log q(\delta_{ij}) =  \delta_{ij}^{(1)} \log\left(\delta_{ij}^{(1)}\right) + \left(1-\delta_{ij}^{(1)}\right) \log\left(1-\delta_{ij}^{(1)}\right),
\end{eqnarray*}
which is further simplified to 
\begin{eqnarray*}
\mathcal{L}_{\bm{\delta},\bm{z}} (\delta_{ij}, z_{ij} \mid \zeta, \bm{\beta}) 
&=& - \frac{1}{2}  \alpha_{ij}^{(2)} + \frac{1}{2}  \left( \alpha_{ij}^{(1)} \right)^2  + \delta_{ij}^{(1)} \log\bigg\{ \Phi\left(\alpha_{ij}^{(1)}\right)\bigg\} + \left( 1 - \delta_{ij}^{(1)}\right) \log\bigg\{ 1 - \Phi  \left(\alpha_{ij}^{(1)}\right) \bigg\} \\
& & - \delta_{ij}^{(1)} \log\left(\delta_{ij}^{(1)}\right) - \left(1-\delta_{ij}^{(1)}\right) \log\left(1-\delta_{ij}^{(1)}\right).
\end{eqnarray*}
In addition, we have
\begin{eqnarray*}
\mathcal{L}_\tau(\tau) & = & \mathbb{E}_{q(\bm{\Omega, \Theta})} \Big\{\log p(\tau) -  \log q(\tau) \Big\}\\
& = & (a_\tau-1) (\log \tau)^{(1)} - b_\tau\tau^{(1)} - \alpha_\tau \log \beta_\tau +\log \Gamma(\alpha_\tau) - (\alpha_\tau-1) (\log \tau)^{(1)} + \beta_\tau \tau^{(1)} \\
& = & (a_\tau-\alpha_\tau) (\log \tau)^{(1)} - (b_\tau - \beta_\tau ) \tau^{(1)} - \alpha_\tau \log \beta_\tau +\log \Gamma(\alpha_\tau),\\
\mathcal{L}_\zeta(\zeta) & = &  \mathbb{E}_{q(\bm{\Omega, \Theta})} \Big\{ \log p(\zeta) - \log q(\zeta)\Big\} \\
& = & - \frac{\zeta^{(2)}}{2 t_0^2} + \frac{n_0 \zeta^{(1)}}{t_0^2} + \frac{1}{2} \Big\{
1 + \log(2\pi\sigma_{\zeta}^2) \Big\},\\
\mathcal{L}_{\bm{\beta}, \bm{\gamma}} (\beta_q, \gamma_q \mid \sigma^2, o) &=& \mathbb{E}_{q(\bm{\Omega, \Theta})} \Big\{\log p(\beta_q \mid \gamma_q, \sigma^2) + \log p(\gamma_q \mid o)-\log q(\beta_q \mid \gamma_q) - \log q(\gamma_q) \Big\}\\ 
& = & \bigg\{\gamma_q\left( \frac{1}{2} \log \sigma^{-2}   - \frac{\beta_q^2}{2} \sigma^{-2}\right)\bigg\}^{(1)} +  \Big\{ \left(1-\gamma_q\right) \log \delta(\beta_q) \Big\}^{(1)} \\ \nonumber
&  & +  \bigg[ \gamma_q^{(1)} \left( \log o \right)^{(1)} + \left(1-\gamma_q^{(1)}\right) \Big\{ \log(1-o) \Big\}^{(1)} \bigg]\\
& & -\mathbb{E}_{q(\bm{\Omega, \Theta})}  \log q(\beta_q \mid \gamma_q)  - \mathbb{E}_{q(\bm{\Omega, \Theta})}  \log q(\gamma_q),
\end{eqnarray*}
with
\begin{eqnarray*}
 \mathbb{E}_{q(\bm{\Omega, \Theta})} \log q(\beta_q \mid \gamma_q)& = & - \frac{1}{2} \gamma_q^{(1)} \bigg\{ 1+\log\left(2\pi\sigma^2_{\beta,q} \right) \bigg\} + \left(1-\gamma_q^{(1)}\right) \bigg\{ \log\delta(\beta_q) \bigg\}^{(1)}, \\
 \mathbb{E}_{q(\bm{\Omega, \Theta})} \log q(\gamma_q) & = &  \gamma_q^{(1)} \log\left(\gamma_q^{(1)}\right) + \left(1-\gamma_q^{(1)}\right) \log\left(1-\gamma_q^{(1)}\right),
\end{eqnarray*}
\begin{eqnarray*}
\mathcal{L}_{o} (o) &=& \mathbb{E}_{q(\bm{\Omega, \Theta})} \Big\{ \log p(o) - \log q(o) \Big\}\\
&=& (a_o - 1) \left( \log o \right)^{(1)} + (b_o-1 ) \Big\{ \log(1-o) \Big\}^{(1)} \\
& & + \log B(\alpha_o, \beta_o) - (\alpha_o - 1) \left( \log o \right)^{(1)} - (\beta_o - 1) \Big\{ \log (1-o) \Big\}^{(1)} \\
& = &(a_o - \alpha_o) \left( \log o \right)^{(1)} + (b_o-\beta_o ) \Big\{ \log(1-o) \Big\}^{(1)} + \log B(\alpha_o, \beta_o) , \\
\mathcal{L}_{\sigma} (\sigma^{-2}) &=&\mathbb{E}_{q(\bm{\Omega, \Theta})} 
\Big\{\log p(\sigma^{-2}) - \log q(\sigma^{-2}) \Big\}\\
& = & (a_\sigma-1) \left(\log \sigma^{-2}\right)^{(1)} - b_\sigma \left(\sigma^{-2} \right)^{(1)} \\
& & - \alpha_\sigma \log \beta_\sigma +\log \Gamma(\alpha_\sigma) - (\alpha_\sigma-1) \left( \log \sigma^{-2} \right)^{(1)} + \beta_\sigma \left( \sigma^{-2}\right)^{(1)} \\
& = & (a_\sigma-\alpha_\sigma) \left(\log \sigma^{-2}\right)^{(1)} - (b_\sigma - \beta_\sigma ) \left(\sigma^{-2}\right)^{(1)} - \alpha_\sigma \log \beta_\sigma +\log \Gamma(\alpha_\sigma),
\end{eqnarray*}
where $B(a,b) = \Gamma(a)\Gamma(b)/\Gamma(a+b)$ is the beta function.

Finally, we simplify the term $\alpha_{ij}^{(2)} - \left(\alpha_{ij}^{(1)}\right)^2 $ in $\mathcal{L}_{\bm{\delta},\bm{z}} (\delta_{ij}, z_{ij} \mid \zeta, \bm{\beta})$, and this is identified as variance $\text{Var}_{q(\bm{\Omega, \Theta})} ( \alpha_{ij} )$. Therefore, 
\begin{eqnarray*}
\alpha_{ij}^{(2)} - \left(\alpha_{ij}^{(1)}\right)^2 & = & 
\text{Var}_{q(\bm{\Omega, \Theta})} (\zeta) + \sum_q V_{iq}^2 \text{Var}_{q(\bm{\Omega, \Theta})} (\beta_q) + \sum_q V_{jq}^2 \text{Var}_{q(\bm{\Omega, \Theta})} (\beta_q), 
\end{eqnarray*}
with
\begin{eqnarray*}
\text{Var}_{q(\bm{\Omega, \Theta})} (\zeta) &= &\sigma_\zeta^{2}, \quad \text{Var}_{q(\bm{\Omega, \Theta})} (\beta_q) = \beta_q^{(2)} - \left( \beta_q^{(1)} \right)^2.
\end{eqnarray*}

\subsection{Hyperparameter settings}\label{app:sparsity}
In this section, we describe the hyperparameter settings for the prior distribution of $\zeta$. 
Under the simplifying assumption that no auxiliary variables are encoded, the prior distribution of edge inclusion given $\zeta$ is 
\begin{eqnarray*}
    \delta_{ij} \mid \zeta \sim \text{Bern}\Big\{\Phi(\zeta)\Big\}.
\end{eqnarray*}
Therefore the prior expectation and variance for the number of edges in the network are
\begin{eqnarray*}
\mathbb{E}\left( \sum_{i<j}\delta_{ij} \right) &=& \mathbb{E}\Bigg\{ \mathbb{E}  \left( \sum_{i<j}\delta_{ij} \mid \zeta \right) \Bigg\} = P \mathbb{E} \Big\{\Phi(\zeta)\Big\}, \\
\mathbb{V}\left( \sum_{i<j}\delta_{ij} \right) & = & \mathbb{V}\Big\{P \Phi(\zeta)\Big\} + \mathbb{E}\bigg[P \Phi(\zeta) \Big\{1-\Phi(\zeta)\Big\}\bigg] \\
& = & \mathbb{E}\Big\{P^2\Phi(\zeta)^2\Big\} + \bigg[ \mathbb{E} \Big\{P\Phi(\zeta)\Big\}\bigg]^2 + \mathbb{E} \Big\{P \Phi(\zeta)\Big\} - \mathbb{E} \Big\{P \Phi(\zeta)^2 \Big\} \\
& = & (P^2 - P)   \mathbb{E} \Big\{\Phi(\zeta)^2\Big\} + P \mathbb{E}\Big\{\Phi(\zeta)\Big\} + P^2 \bigg[E\Big\{\Phi(\zeta)\Big\}\bigg]^2.
\end{eqnarray*}
Next, using $\zeta \sim \mathcal{N}(n_0, t_0^2)$, we find
\begin{eqnarray*}
E\bigg\{\Phi(\zeta)\bigg\} &=& \Phi\left(\frac{n_0}{\sqrt{1+t_0^2}}\right),\\
E\bigg\{\Phi(\zeta)^2\bigg\} &=& \Phi\left(\frac{n_0}{\sqrt{1+t_0^2}}\right) - 2T\left(\frac{n_0}{\sqrt{1+t_0^2}},\frac{1}{\sqrt{1+t_0^2}}\right),
\end{eqnarray*}
where $T(h,a)$ is Owen’s T function \citep{owen1956tables},
\begin{eqnarray*}
T(h,a) &=& \varphi(h) \int_0^a \frac{\varphi(hx)}{1+x^2} dx, \quad a, h \in \mathbb{R},
\end{eqnarray*}
with $\varphi(\cdot)$ the standard normal density function and $\Phi(\cdot)$ the standard normal cumulative density function. We thus can proceed by specifying an expectation and a variance for the prior number of edges in the network, and solving for $n_0$ and $t_0^2$.

\begin{figure}[t!]
    \centering
    \includegraphics[width=0.8\textwidth]{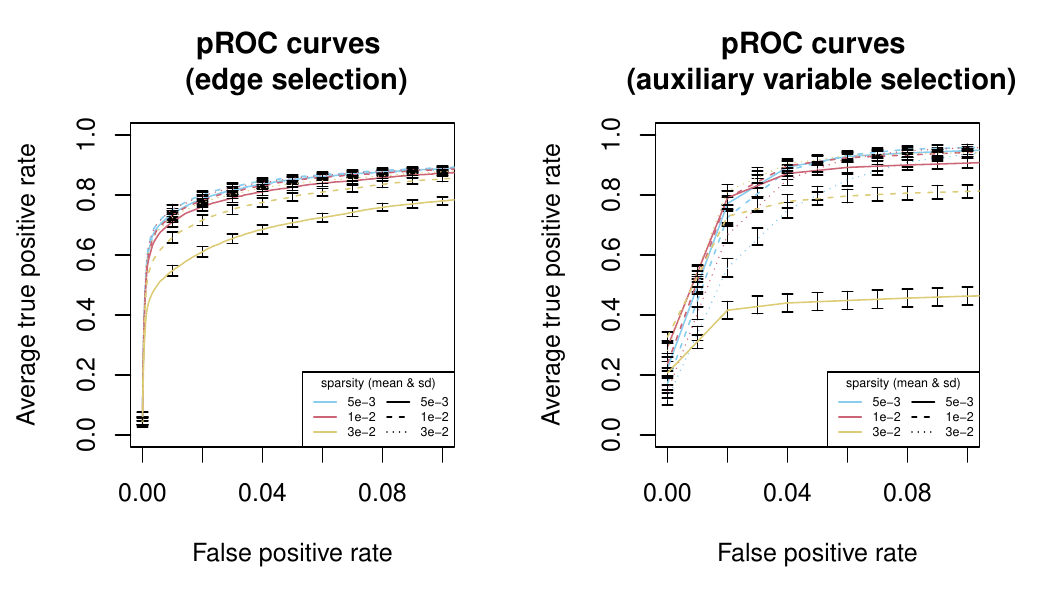}
    \\
    \includegraphics[width=0.9\textwidth]{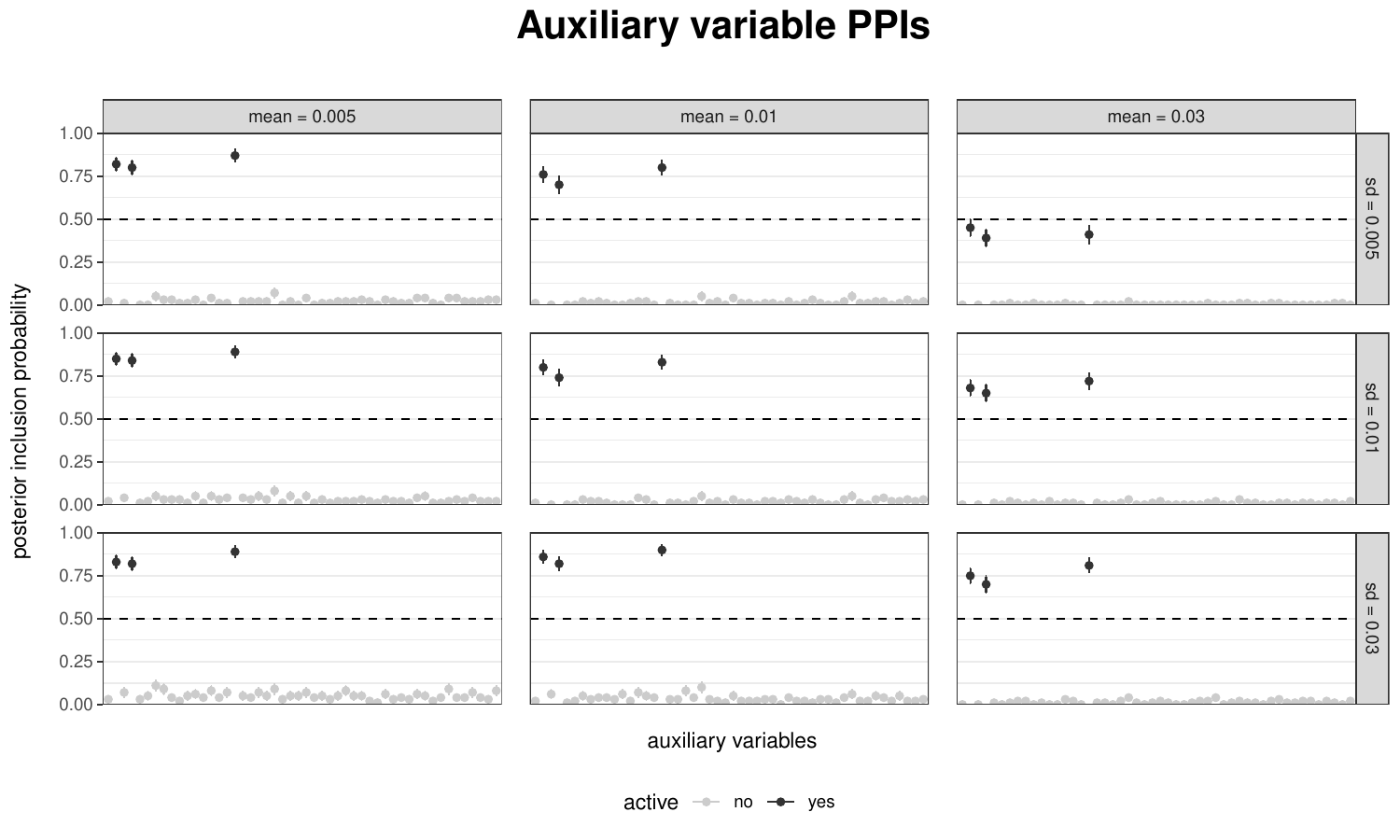}
    \caption{\footnotesize \normalfont Sensitivity to the hyperparameter choices for $\zeta$ using GMSS on simulated data under the ``reference'' scenario. Top left: average pROC curves for edge selection; top right: average pROC curves for auxiliary variable selection; bottom: auxiliary variable PPIs under the hyperparameter settings of Table~\ref{tab:n0t0}, where columns and rows correspond to different prior choices for the mean and standard deviation of the network sparsity. Error bars correspond to standard errors based on 100 replicates.}
    \label{fig:proc_n0t0}
\end{figure}

To assess the sensitivity of this hyperprior specification, we use the ``reference'' simulation scenario, i.e., with $N = 200$, $P= 100$, $Q=50$, $Q_0 = 3$. For this scenario, the average network sparsity is $\approx 3\%$, therefore the sparsity in the absence of variable-triggered hubs should be smaller. We thus evaluate the performance using a prior expectation of sparsity to 0.5\%, 1\% and 3\%. In addition, we consider varying levels of uncertainty and set the prior standard deviation to 0.5\%, 1\% and 3\% 
of the total number of possible edges. 
Table \ref{tab:n0t0} summarises the hyperparameter settings considered and Figure~\ref{fig:proc_n0t0} displays corresponding average partial receiver operating characteristic (pROC) curves for the edge selection and auxiliary-variable selection using GMSS. The performance is comparable across all the hyperparameter settings. The prior expectation of 3\% leads to slightly inferior performance for edge selection but large standard deviations tend to help in this scenario. 
In addition, active auxiliary variables are almost always effectively singled out, except in cases where the prior expectation is set too high and the standard deviation is set too low.
This sensitivity study reassuringly suggests that inference is not very sensitive to reasonable guesses for these hyperparameters. In the simulations presented in the main text, we set the prior expectation to 1\% and the standard deviation to 3\% throughout. 
\begin{table}[!ht]
    \centering
      \begin{tabular}{c c c c c c}
    \hline
    \multicolumn{2}{c}{Prior expectation} & \multicolumn{2}{c}{Prior standard deviation} & \multicolumn{2}{c}{Hyperparameters} \\
    \hline
    \% & number & \% & number & $n_0$ & $t_0^2$\\
    \hline
0.5\% & 25 & 0.5\% & 25 & -2.69 & 0.09\\
0.5\% & 25 & 1\% & 50 & -2.93 & 0.30 \\
0.5\% & 25 & 3\% & 150 & -4.34 & 1.85 \\
1\% & 50 & 0.5\% & 25 & -2.36 & 0.03 \\
1\% & 50 & 1\% & 50 & -2.45 &0.12 \\
1\% & 50 & 3\% & 150 & -3.09 &0.77 \\
3\% & 150 & 0.5\% & 25 & -1.88 & 0.004 \\
3\% & 150 & 1\% & 50 & -1.90 & 0.02 \\
3\% & 150 & 3\% & 150 & -2.04& 0.18 \\
    \hline
    \end{tabular}
    \caption{\footnotesize \normalfont Prior specification for the sparsity parameter $\zeta$, with the hyperparameters $n_0$ and $t_0^2$ derived according to the procedure presented above.
    }
    \label{tab:n0t0}
\end{table}

\subsection{Parallel grid search procedure for spike-and-slab variances}\label{app:ssvar}

In this section, we detail the use of model selection criteria in the grid search procedure to set the spike variance in the edge-selection bottom-level model. We compare the Akaike information criterion (AIC) adopted in the main text with the Bayesian information criterion (BIC) and the extended Bayesian information criterion (EBIC) on $100$ replicates of the ``reference'' data generation scenario, namely, with $N = 200$ samples, $P= 100$ nodes and $Q=50$ auxiliary variables, of which $Q_0 = 3$ contribute to the node degrees. AIC,  BIC and EBIC are defined as follows
\begin{eqnarray*}
\text{AIC}(\nu_0) & = &
- N \log|\hat{\bm \Omega}^*| + \text{tr}(\bm{Y}^T \bm{Y}\hat{\bm \Omega}^*) + 2 \sum_{i<j} \mathds{1}\Bigg\{\hat{\delta}_{ij}^{(1)} \geq 0.5 \Bigg\}, \\ \nonumber
\text{BIC}(\nu_0) & = &
- N \log|\hat{\bm \Omega}^*| + \text{tr}(\bm{Y}^T \bm{Y}\hat{\bm \Omega}^*) +  \log(N) \sum_{i<j} \mathds{1}\Bigg\{\hat{\delta}_{ij}^{(1)} \geq 0.5 \Bigg\}, \\ \nonumber
\text{EBIC}(\nu_0) & = & \text{BIC}(\nu_0) + 4\gamma \log(P) \sum_{i<j} \mathds{1}\Bigg\{\hat{\delta}_{ij}^{(1)} \geq 0.5 \Bigg\},
\end{eqnarray*}
where $\hat{\bm \Omega}$ and $\hat{\delta}_{ij}^{(1)}$ represent the posterior estimates of precision matrix and edge $(i,j)$'s inclusion probability, $\hat{\bm \Omega}^*$ refers to a thresholded precision matrix defined by

\begin{eqnarray*}
\hat{\omega}_{ij}^* &=&
\begin{cases}
\hat{\omega}_{ij}, \quad &\text{if } i = j, \\
\hat{\omega}_{ij}, \quad &\text{if } \hat{\delta}_{ij}^{(1)} \geq 0.5 \text{ and } i \neq j,\\
0, \quad &\text{if } \hat{\delta}_{ij}^{(1)} < 0.5 \text{ and } i \neq j,\\  
\end{cases}
\end{eqnarray*}
and $\gamma$ is a tuning parameter of EBIC with default choice of $0.5$ \citep{chen2008extended}.

\begin{figure}[ht!]
\raggedright
\hspace{0.3cm}\includegraphics[width=0.66\textwidth]{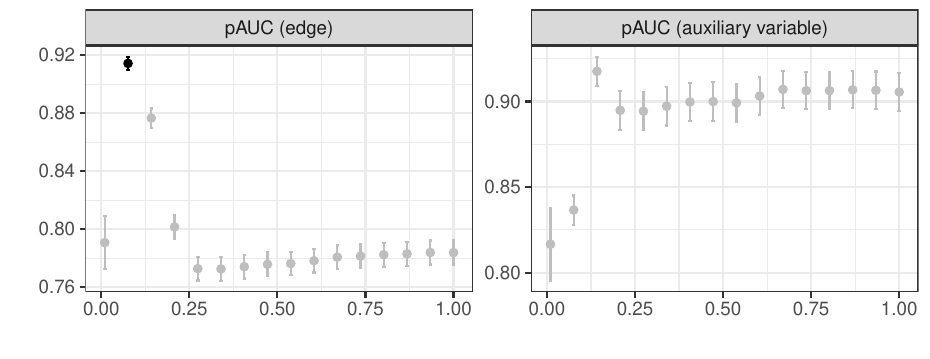}
\includegraphics[width = \textwidth]{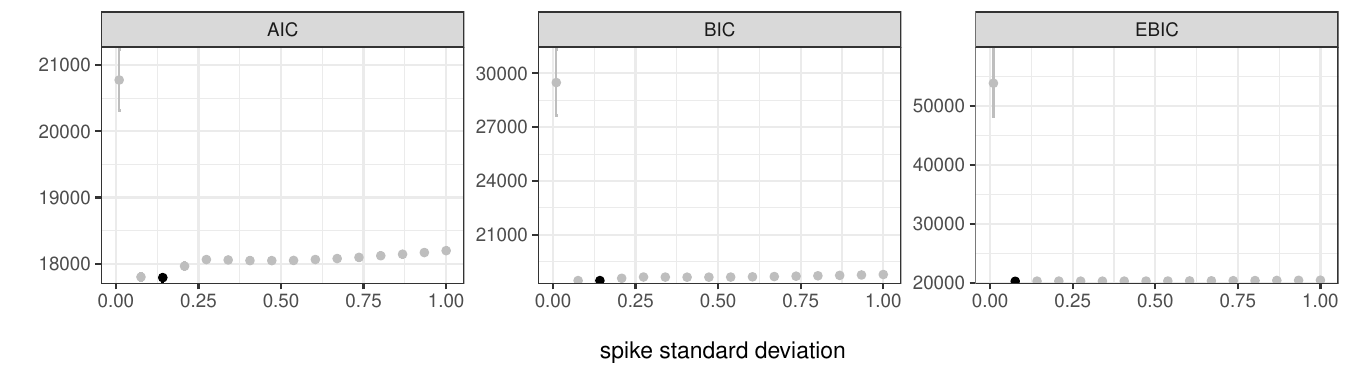}
\caption{\footnotesize \normalfont Spike standard deviation selection. The average partial area under the curve (pAUC) for edge and auxiliary variable selection and model selection criteria (AIC,  BIC and EBIC), based on $100$ replicates, are shown, for a grid of spike standard deviations $\nu_0$. In each case, the best model, with the lowest model selection criteria or highest average pAUC for edge selection, is highlighted in black. Error bars represent standard errors.}
\label{fig:model_criterion}
\end{figure}

Figure~\ref{fig:model_criterion} compares the model selection criteria with the resulting standardised pAUCs for edge selection and auxiliary variable selection performance, for a grid of spike standard deviations. The best average pAUC is achieved at $\nu_0 = 0.07$ for edge selection, which corresponds to the lowest values of all three model selection criteria.
The performance measures stabilise and remain satisfactory (pAUC $> 0.8$) for variable selection for $\nu_0 \geq 0.07$, while the edge selection performance deteriorates quickly after $\nu_0 = 0.07$. The values of the model selection criteria do not vary much after the smallest spike standard deviation, except for a slight upward trend in AIC after $\nu_0 = 0.07$. The three criteria point to similar models; we implemented all of them in our R package, with the AIC as a default choice.

\subsection{Bayesian false discovery rate}\label{sec:app:bfdr}
Given posterior probabilities of inclusion, $\delta_{ij}^{(1)} = \text{pr}\left(\delta_{ij} = 1 \mid \bm Y\right)$, for edges $(i,j)$, $1 \leq i<j \leq P$, and a pre-specified threshold $\kappa\in [0,1]$, the corresponding Bayesian false discovery rate (FDR) can be estimated as 
\begin{eqnarray}\label{eq:fdr}
\widehat{\text{FDR}}(\kappa) = \frac{\sum_{i<j} \left(1 - \delta_{ij}^{(1)}\right)\mathds{1}\left\{\delta_{ij}^{(1)} > \kappa\right\}}{\sum_{i<j}\mathds{1}\left\{\delta_{ij}^{(1)} > \kappa\right\}}\,,
\end{eqnarray}
following \cite{newton2004detecting}. One can obtain a threshold corresponding to a target Bayesian FDR by interpolating a series of FDR estimates computed for a grid of thresholds $0 < \kappa_1 < \cdots < \kappa_T < 1$.

\FloatBarrier
\newpage

\section{Addendum to the simulation experiments}
\subsection{Variants of the model}\label{sec:app:sim_details:formula}
To ensure comparability between the GM model and the GMN and GMSS models, we introduced a variant of GM, hereafter called GM$^*$, which replaces the beta prior on the edge inclusion probabilities by a normal prior within probit link, that is, \eqref{eq:gmrho} in the main text is replaced by 
\begin{eqnarray*}
\delta_{ij} \mid \rho &\sim& \text{Bern}(\rho),\\ \nonumber
\Phi^{-1}(\rho) &\sim& \mathcal{N}(n_0,t_0^2). 
\end{eqnarray*}
This allows us to use the same hyperprior specification procedure (Supplementary Material~\ref{app:sparsity}), and hence the same $n_0$ and $t_0^2$ for all three models, therefore putting them all on an equal footing for a fair comparison in Section~\ref{sec:sim:accuracy} of the main text.

\subsection{Edge-selection performance} \label{sec:app:sim:ref_gm_gmss}
\begin{figure}[ht!]
\centering
\includegraphics[width = 0.85\textwidth]{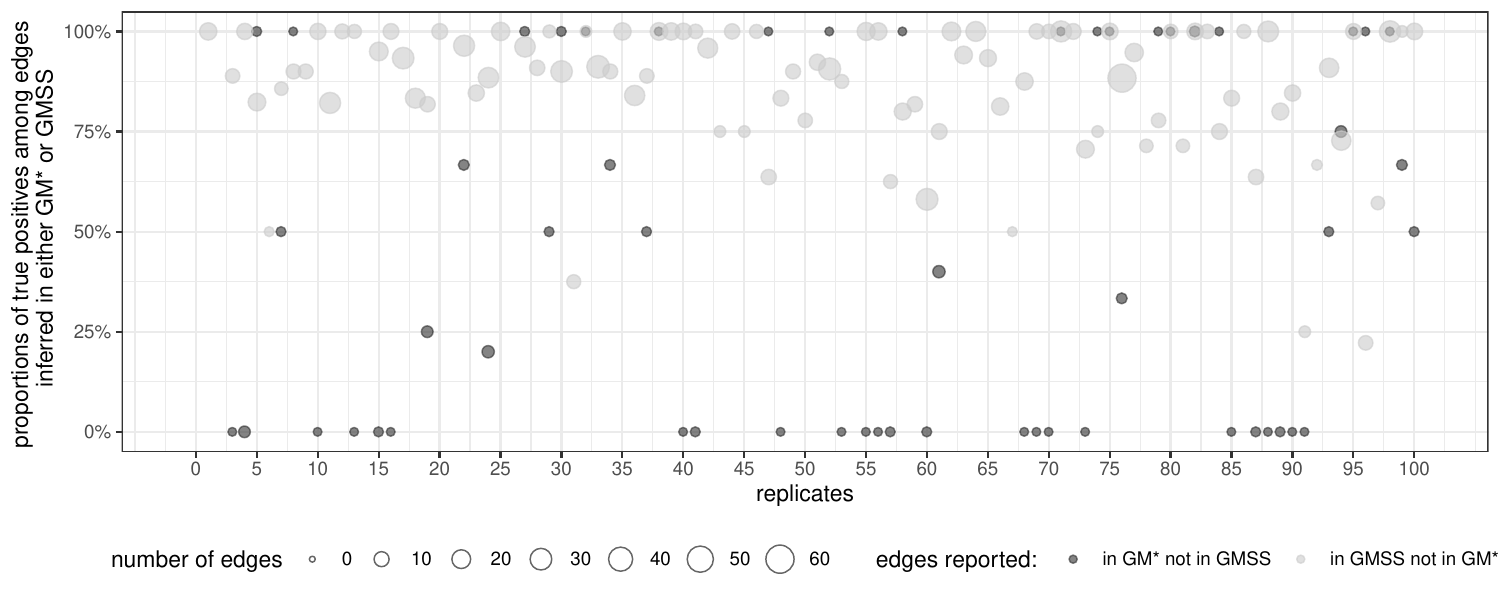}
\caption{\footnotesize \normalfont  Proportion of true positives among edges reported by GMSS but not by GM$^*$ (light grey), and vice versa (dark grey) for each of the $100$ data replicates ($x$-axis) of the reference simulation scenario. The size of points represents the number of edges detected exclusively by the corresponding method. Note that, for a given replicate, GMSS (resp. GM$^*$) may not report any edge that GM$^*$ (resp. GMSS) doesn't report, in which case, no point will appear.}
\label{fig:unique_edges_supp}
\end{figure}

We complement our illustration of the improved edge-selection performance of GMSS, compared with GM$^*$, in the reference simulation scenario
(the left panel in Figure~\ref{fig:small_ref_inference}B and Section~\ref{sec:sim:accuracy} in the main text). 
As expected, inspecting the edges reported by GMSS but that GM$^*$ failed to detect indicates that they correspond to edges between nodes influenced by the selected auxiliary variables.
Figure~\ref{fig:unique_edges_supp} reports the proportion of true positives among the edges reported by GMSS but not by GM$^*$ (light grey), and vice versa (dark grey) for each of the $100$ data replicates. We observe that GMSS tends to detect more edges, a substantial proportion of which are true edges. In contrast, GM$^*$ generally selects fewer than 10 edges not selected by GMSS, and these are often false positives. Moreover, the posterior inclusion probabilities of edges reported by GMSS but not by GM$^*$ (light grey) are all $<0.1$ in the GM$^*$ estimation but $>0.9$ in the GMSS estimation.

\subsection{Non-positive effects}\label{sec:app:nonpositive}
The numerical experiments presented in the main text (Section~\ref{sec:simulation}) focus on problems where auxiliary variables are associated with the \emph{presence} of hubs, that is, thus, edges, i.e., 
positive effects $\bm\beta$ of auxiliary variables on the propensity of nodes to have high degrees. In this section, we investigate the performance of GMSS when auxiliary variables have a repressing effect on the propensity of nodes to have high degrees (negative $\bm\beta$, referred to as ``negative scenario''), or in presence of both hub-inducing for some auxiliary variables \emph{and} hub-repressing effects for other auxiliary variables (referred as ``combined scenario'').

We set $\zeta$ such that network sparsity is around 3\% in both the negative and combined scenarios, with $Q_0 = 3$ and $2$ active variables, respectively. In the combined scenario, the first active auxiliary variable has a positive effect and the second has a negative effect. 
We generate adjacency matrices, precision matrices and data following the procedure in Section~\ref{sec:simulation} of the main text and for $100$ data replicates.

\begin{figure}[ht!]
\centering
\includegraphics[width=0.42\textwidth]{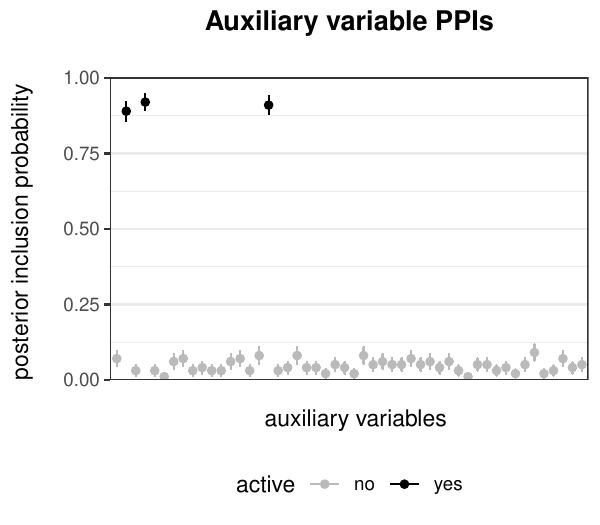}
\includegraphics[width=0.42\textwidth]{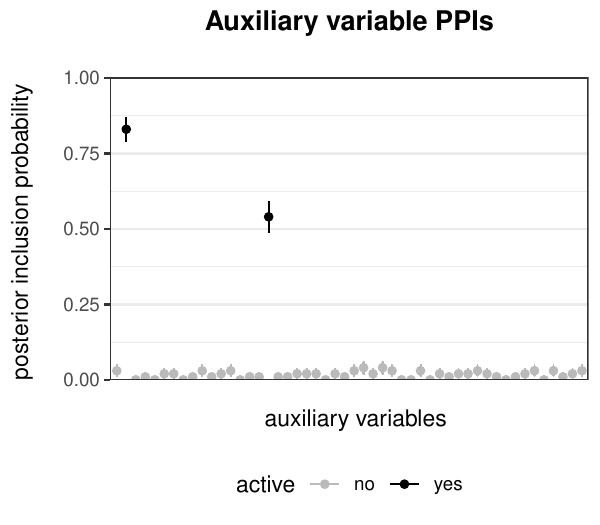}
\caption{\footnotesize \normalfont  Performance under scenarios when node-level variables have negative effects ($Q_0=3$; left) and a combination of positive and negative effects ($Q_0=2$; right), with $N=200$ samples, $P = 100$ nodes, $Q = 50$ candidate auxiliary variables and a network sparsity of about 3\%.}
\label{fig:opposite_supp}
\end{figure}

We then apply GMSS to estimate graph structures and effects of auxiliary variables.  Figure~\ref{fig:opposite_supp} indicates that GMSS is able to disentangle the active variables from the inactive ones in both scenarios.
The average PPI corresponding to the negative effect of the combined scenario is somewhat lower, likely because the detection of the negative signals is being offset by the level of overall sparsity assumed. 

\subsection{Null scenario}\label{sec:app:sim:null}
\begin{figure}[ht!]
\centering
\includegraphics[width=0.34\textwidth]{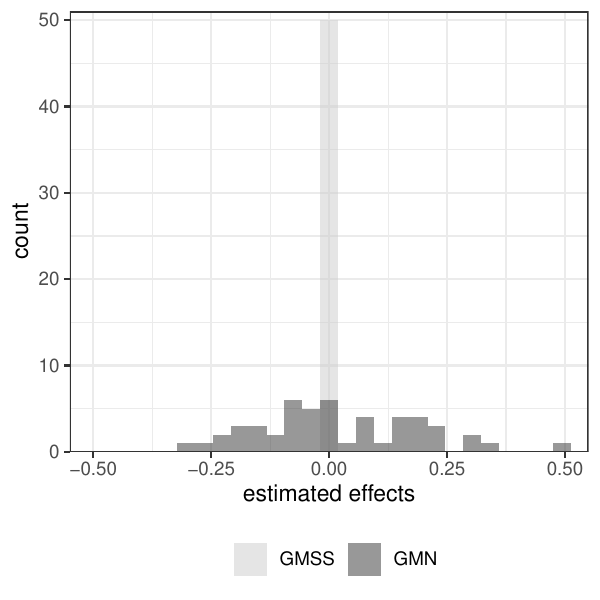} \;\includegraphics[width=0.45\textwidth]{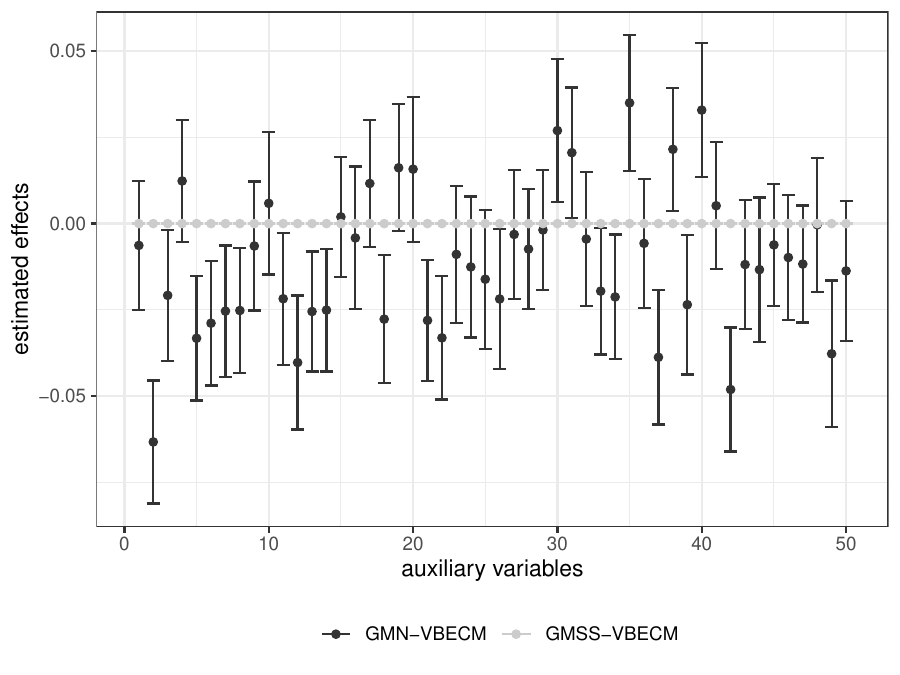}
\caption{\footnotesize \normalfont Auxiliary variable selection under a null-model scenario, with no active auxiliary variable. Left: histograms for effects of $Q=50$ candidate auxiliary variables, estimated from GMN and GMSS approaches, for the first data replicate in a problem with $N = 200$ samples and $P = 100$ nodes; none of the auxiliary variables was used to simulate the graph structure.  Right: average effects estimated using GMSS (grey) and GMN (black) across all $100$ replicates, with standard error bars. The GMSS average effects are essentially zero, with standard error too small to be visible.}
\label{fig:q50q00_gmss}
\end{figure}

Figure~\ref{fig:q50q00_gmss} presents the auxiliary variable effects estimated by GMSS and GMN, under the null-model scenario discussed in Section~\ref{sec:simulation:null} of the main text, for the first replicate (left) and averaged over $100$ replicates (right). 

\subsection{A misspecified scenario: similarity-based edge effects}\label{sec:app:misedge}
In this section, we consider a simulation study where the edge model is misspecified, i.e., the assumed submodel on the edge inclusion parameter, $$\delta_{ij} \mid \zeta, \bm \beta \sim \text{Bern}\left\{\Phi(\zeta + \sum_{q} V_{iq} \beta_q + \sum_q V_{jq} \beta_q ) \right\},$$ does not reflect the data-generation mechanism.
Specifically, we simulate information such that the similarity between auxiliary data for any pair of nodes influences the presence or absence of edges between the nodes. Such a setting may, for instance, be relevant in the context of brain connectivity networks, whereby nodes – quantifying fMRI signal in regions of interest – are more likely to share edges with nodes from nearby brain regions \citep{bu2021integrating}. In this case, information about the location of these nodes may be used as auxiliary information in a similarity-based edge submodel, such as
\begin{equation}\label{eq:misedge}
\delta_{ij} \mid \zeta, \bm \beta \sim \text{Bern}\left[\Phi\{\zeta + \sum_{q} \exp (-|V_{iq} - V_{jq}|) \beta_q \} \right].
\end{equation}

To assess the behaviour of GMSS for such a misspecified setting, we produce $100$ data replicates, as follows: we generate $Q=50$ candidate auxiliary variables, of which  
$2$ influence the inclusion of edges, assuming the above similarity-based edge model~\eqref{eq:misedge}. Specifically, we use $\beta_q \neq 0$ for $q \in \{2, 44\}$, and $\beta_q = 0$ for $q\notin \{2,44\}$. We generate the ``inactive'' auxiliary variables from independent standard normal distributions, and the two ``active'' variables from three bivariate normal distributions with different mean vectors, to reflect the fact that the nodes can be ``grouped'' into three distinct ``regions'' of a two-dimensional space. Specifically, we use the mean vectors $(-1,0)$, $(0,1)$, and $(1,0)$ for 30, 30 and 40 nodes, respectively. We then simulate $N=200$ observations for $P = 100$ nodes following Section~\ref{sec:simulation:setup} in the main text. This procedure implies that edges are more likely between nodes located in the same ``region of the space'', i.e., with \emph{similar values} of the active variables. GMSS applied to this data selects neither of the active variables (PPIs $< 0.1$). In other words, although the similarity between the two auxiliary features influences the edge pattern, the features themselves don't influence the propensity of nodes to have high degree, hence GMSS appropriately discards them as irrelevant to the centrality of nodes.

\subsection{Handling posterior multimodality}\label{sec:app:multimod}
Figure~\ref{fig:small_ref_inference}C and Section~\ref{sec:sim:comparison} of the main text illustrate the improved performance achieved by GMSS-VBECM, compared with GMSS-ECM. In this section, we investigate the possible reasons for this performance gap. 

\begin{figure}[ht!]
\centering
\includegraphics[width = 0.8\textwidth]{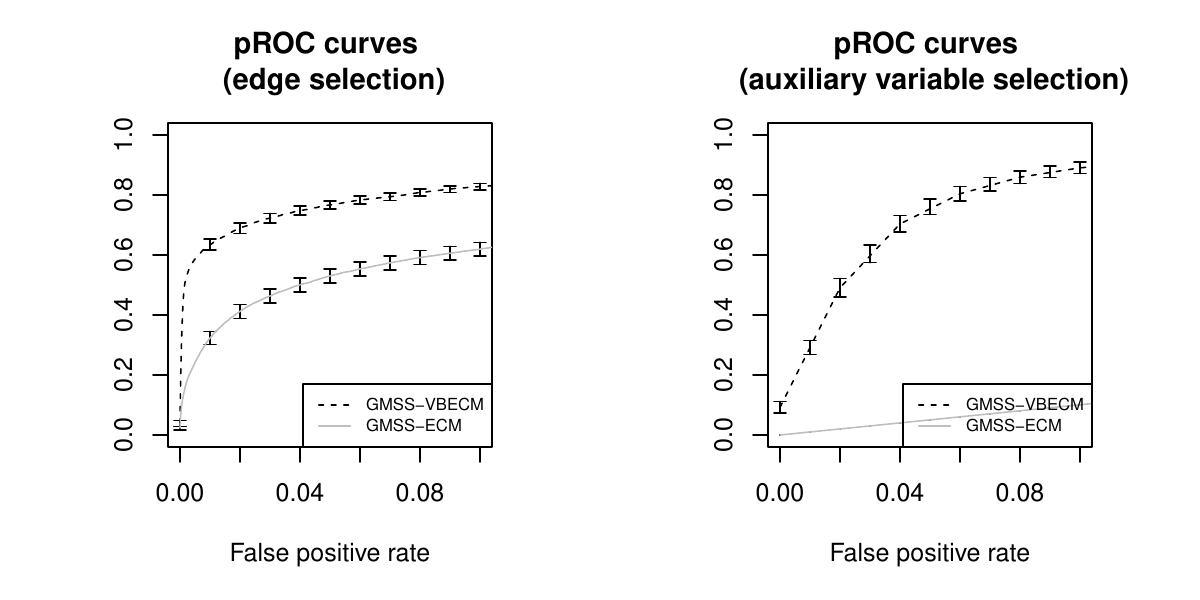}
\caption{\footnotesize \normalfont Comparison between VBECM and ECM inference for the GMSS model for a given spike-and-slab prior configuration. Average pROC curves for edge selection (left) and auxiliary variable selection (right) using GMSS, along with standard error bars based on $100$ replicates. We consider spike-and-slab priors with $\nu_1 = 100, \nu_0 = 0.07, \sigma_1=1, \sigma_0 = 10^{-6}$.}
\label{fig:vbem_em_1ss}
\end{figure}

We first examine whether the performance of the ECM approach is impacted by the need for a second grid search procedure for spike-and-slab variances at the level of the auxiliary variable effects.
To assess this, we consider the following top-level spike-and-slab prior formulations:
\begin{eqnarray*} 
\beta_q \mid \gamma_q, \sigma^2 &\sim&  \gamma_{q}\mathcal{N}(0,\sigma^2) + (1-\gamma_{q}) \delta(\beta_q) , \quad q=1,\ldots, Q,\\ \nonumber
\nonumber
\sigma^{-2} &\sim& \text{Gamma}(a_\sigma, b_\sigma),
\end{eqnarray*}
used in the VBECM approach, and 
\begin{eqnarray*} 
\beta_q \mid \gamma_q, \tau_2 &\sim&  \gamma_{q}\mathcal{N}(0,\tau_2^{-1}) + (1-\gamma_{q}) \mathcal{N}(0,\sigma_0^2 \tau_2^{-1}) , \quad q=1,\ldots, Q,\\ \nonumber
\nonumber
\tau_2 &\sim& \text{Gamma}(a_\sigma, b_\sigma),
\end{eqnarray*}
used in the ECM approach, i.e., setting $\sigma_1 = 1$ in Equation~(\ref{eq:app:css_var}) of the Supplementary Material~\ref{app:emdev}. We further make the two formulations as comparable as possible by setting $\sigma_0$ to a small value, $10^{-6}$, to mimic a discrete spike. Finally, we set the hyperparameters in the top-level spike-and-slab (Equation~\eqref{eq:cssprior} in the main text) to be $\nu_1 = 100$ and $\nu_0 = 0.07$, since these choices achieve the lowest average AIC in our simulations (Figure~\ref{fig:model_criterion}). Despite these adjustments, Figure~\ref{fig:vbem_em_1ss} below still indicates a significant performance gap between VBECM and ECM, suggesting that the performance gap may not arise from the ``double grid search'' nor a discrete formulation of spike-and-slab, but rather from the inference algorithms themselves. 

We next investigate the potential ability of variational inference to mitigate entrapment in local modes. We expect that this may result from approximating of full posterior distributions, unlike with the ECM algorithm. Specifically, using the structured mean-field approximation of Equation~\eqref{eq:mf} of the main text retains a joint distribution for the spike-and-slab parameters $\beta_q$ and $\gamma_q$. To explore this, we consider one data replicate and run the two inference algorithms using 200 different random starts. 
Figure~\ref{fig:small_ref_inference}C of the main text
displays the obtained optimal values for the objective functions ($Q$ function for the ECM algorithm and ELBO for the VBECM algorithm): the $Q$ function values obtained by the ECM algorithm exhibit high variability, suggesting that it reaches different local optima. In contrast, the ELBO values reached by the VBECM algorithm are consistently high across the random starts.

\subsection{Variational credible intervals}\label{sec:app:uq}
In this section, we illustrate the advantages of  quantifying uncertainty around the estimates using variational inference for the GMSS model. 
We simulate $N=100$ independent samples from a $P=20$ dimensional multivariate normal distribution, whose precision matrix structure is influenced by one auxiliary variable out of 10 candidate variables. We set the overall network sparsity to be 1\%, i.e., $\zeta = \Phi^{-1}(0.01)$. The effect of the fifth variable is set to 2, and all other effects are set to 0. For each replicate, we use the procedure described in Section~\ref{sec:simulation:setup} of the main text to generate auxiliary variables and simulate the adjacency matrix such that $A_{ij} \sim \text{Bern}\left\{\Phi(\zeta + 2V_{i5} +2V_{j5})\right\}$, for $1\leq i<j \leq P$. The precision matrix and data are generated as described in Section~\ref{sec:simulation:setup} of the main text.
\begin{figure}[ht!]
\centering
\includegraphics[width = \textwidth]{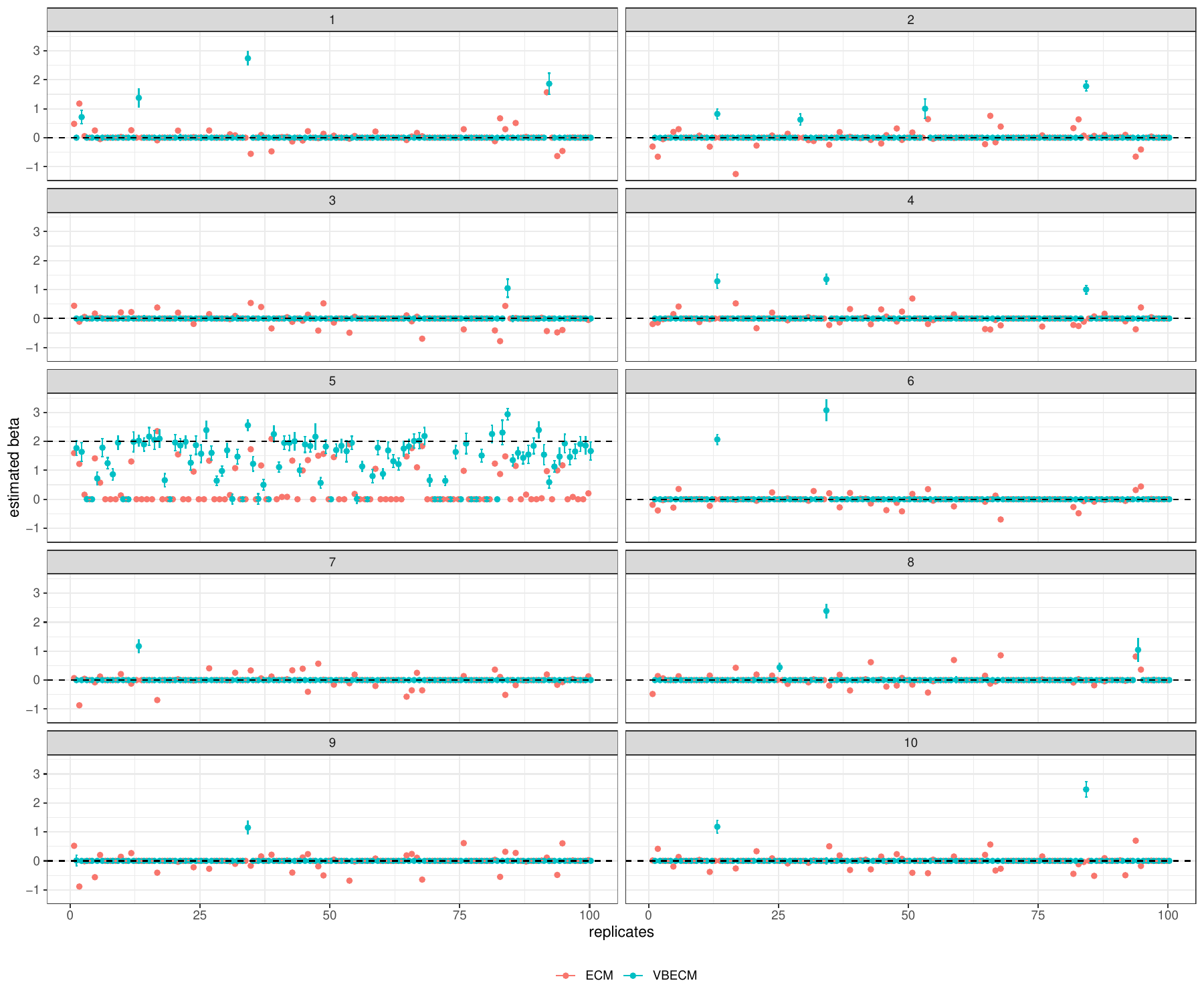}
\caption{\footnotesize \normalfont Auxiliary effect estimates by GMSS-ECM (red) and GMSS-VBECM (blue) in an example with $N=100$ samples, $P=20$ nodes and $Q = 10$ candidate auxiliary variables. Each panel represents one auxiliary variable and the true, simulated effects are highlighted by the dashed horizontal lines. The estimated effects using GMSS-ECM and GMSS-VBECM are shown for 100 replicates ($x$-axis). Error bars for the VBECM algorithm correspond to the 95\% variational credible intervals (unavailable for the ECM algorithm).}
\label{fig:emvbem_uq}
\end{figure}

We compare the estimations of the effects of auxiliary variables obtained with GMSS-ECM and GMSS-VBECM. 
Figure~\ref{fig:emvbem_uq} shows that the ECM algorithm does not always produce null estimates for the inactive auxiliary variables (i.e., all but the fifth), while the variational credibility bars produced by the VBECM algorithm cover zero in more than $95\%$ of the cases. For the active auxiliary variable (the fifth variable), the VBECM algorithm also tends to provide better estimates compared to the ECM algorithm, although the coverage of the variational credible intervals sometimes fail to cover the true value. This is likely because of the tendency of variational inference to underestimate posterior variances. This is a well-known drawback resulting from optimising the \emph{reverse} Kullback--Leibler divergence (which tends to prevent the approximation from putting mass in region of the parameter space where the true distribution has little mass) as well as from the use of factorised mean-field distributions. The sparse prior specification, here the spike-and-slab prior, likely results in further unwanted shrinkage towards zero. However, it is important to note that our VBECM algorithm clearly outperforms the ECM implementation for variable selection, which, rather than effect size estimation, remains our primary goal.

\subsection{Runtime profiling }\label{sec:app:sim:runtime}
We next present the runtime profiling for all simulation scenarios of Table~\ref{tab:sim:sensitivity} of the main text. Table~\ref{tab:sim:all:time} indicates that, although the GM approach is the fastest due to its simpler model with no encoding of auxiliary variables,  the top-level spike-and-slab specification of the GMSS approach permits saving about two-third of the computational time used by the GMN approach. The GMSS runs complete in less than or about one minute on average, across all the data scenarios. 

\begin{table}[!ht]
\centering
\resizebox{0.9\columnwidth}{!}{
\begin{tabular}{|c| c| c|c|c|c|c|ccc|} 
\hline
& Sparsity & Noise   & $Q_0$ & $Q$ & $N$ & $P$  &  \multicolumn{1}{c|}{GM$^*$} & \multicolumn{1}{c|}{GMN} & \multicolumn{1}{c|}{GMSS}\\
\hline
1. & \cellcolor{gray!25} & 
\cellcolor{gray!25}
&  
\cellcolor{gray!25}
& \cellcolor{gray!25} & \cellcolor{gray!25} &\multirow{-1}{*}{ \cellcolor{gray!25}100} & 3.56 (0.17)  & 63.63 (1.92)  & 41.69 (1.27)  \\
\hhline{~~~~~~----}
2. & \cellcolor{gray!25}& \cellcolor{gray!25}& \cellcolor{gray!25}&  \cellcolor{gray!25} &\multirow{-2}{*}{\cellcolor{gray!25}200} & 50 & 0.48 (0.02)  & 4.97 (0.12)  & 5.02 (0.24) \\
\hhline{~~~~~-----}
3. & \cellcolor{gray!25}& \cellcolor{gray!25}& \cellcolor{gray!25}& \cellcolor{gray!25} & \multirow{2}{*}{100} & \cellcolor{gray!25} 100 &  14.66 (0.21)  & 167.71 (6.42)  & 51.94 (1.06)  \\
\hhline{~~~~~~----}
4. & \cellcolor{gray!25}& \cellcolor{gray!25}& \cellcolor{gray!25}& \multirow{-4}{*}{\cellcolor{gray!25}50} & & 50 & 0.84 (0.01)  & 5.92 (0.10)  & 5.69 (0.15)  \\
\hhline{~~~~------}
5. & \cellcolor{gray!25}& \cellcolor{gray!25} & \cellcolor{gray!25} & 20 & \cellcolor{gray!25}  &  \cellcolor{gray!25} & 3.40 (0.17)  & 35.09 (0.78)  & 22.53 (0.64)  \\
\hhline{~~~~-~~---}
6. & \cellcolor{gray!25}&\cellcolor{gray!25} & \multirow{-6}{*}{\cellcolor{gray!25}3}& 100 & \cellcolor{gray!25} &\cellcolor{gray!25} & 3.51 (0.17)  & 73.71 (2.73)  & 60.73 (1.63)  \\
\hhline{~~~--~~---}
7. & \cellcolor{gray!25}& \cellcolor{gray!25}&  1 & \cellcolor{gray!25} & \cellcolor{gray!25} & \cellcolor{gray!25} &  2.96 (0.09)  & 57.75 (1.36)  & 52.17 (1.23)  \\
\hhline{~~~-~~~---}
8. & \cellcolor{gray!25}& \multirow{-8}{*}{\cellcolor{gray!25}10\%}& 5 & \cellcolor{gray!25} & \cellcolor{gray!25} & \cellcolor{gray!25} & 3.35 (0.17)  & 61.07 (1.75)  & 30.30 (0.83)  \\
\hhline{~~--~~~---}
9. & \cellcolor{gray!25}& 20\% & \cellcolor{gray!25} &  \cellcolor{gray!25} & \cellcolor{gray!25} & \cellcolor{gray!25} & 3.44 (0.17)  & 57.38 (2.09)  & 35.53 (1.08)  \\
\hhline{~~-~~~~---}
10. & \multirow{-10}{*}{\cellcolor{gray!25}3\%} & 30\% & \cellcolor{gray!25} & \cellcolor{gray!25} & \cellcolor{gray!25} & \cellcolor{gray!25} & 3.26 (0.15)  & 56.78 (1.86)  & 35.47 (0.88)  \\
\hhline{~--~~~~---}
11. & 1\% &  \cellcolor{gray!25} & \cellcolor{gray!25}  & \cellcolor{gray!25} & \cellcolor{gray!25} & \cellcolor{gray!25} &  3.13 (0.16)  & 48.28 (1.41)  & 32.80 (1.07) \\ 
\hhline{~-~~~~~---}
12. & 10\% & \multirow{-2}{*}{\cellcolor{gray!25}8.5\%} & \multirow{-4}{*}{\cellcolor{gray!25}3} & \multirow{-6}{*}{\cellcolor{gray!25}50} & \multirow{-6}{*}{\cellcolor{gray!25}200}& \multirow{-8}{*}{\cellcolor{gray!25}100} & 6.07 (0.22)  & 69.81 (2.76)  & 29.47 (0.82)  \\
\hline
\end{tabular}
}
\caption{\footnotesize \normalfont Average runtime, including grid search, in seconds for GM$^*$, GMN and GMSS in the simulation experiments presented in 
Table~\ref{tab:sim:sensitivity} (main text) on an Intel Xeon CPU, 2.60 GHz.
Standard errors based on 100 replicates are in parentheses.} 
\label{tab:sim:all:time}
\end{table}

\clearpage
\newpage
\FloatBarrier

\section{Addendum to the monocyte network application} \label{sec:app:application}
Table~\ref{tab:hubgenes} presents hub genes in the monocyte networks inferred by both GM$^*$ and GMSS methods.
Table~\ref{tab:literaturegenes} and Table~\ref{tab:literaturegenes_stimulated} list the neighbours of the \emph{LYZ}, \emph{YEATS4} and \emph{CREB1} identified using the GM$^*$ and GMSS approaches, for unstimulated and stimulated monocyte data respectively.

\begin{table}[ht!]
\centering
\resizebox{\columnwidth}{!}{%
\begin{tabular}{cc ccc|cc ccc}
\hline
\multicolumn{5}{c|}{unstimulated} & \multicolumn{5}{c}{stimulated} \\
\multicolumn{2}{c}{GM$^*$} & \multicolumn{3}{c|}{GMSS} &\multicolumn{2}{c}{GM$^*$} & \multicolumn{3}{c}{GMSS} \\
Gene & Degree & Gene & Degree & Change & 
Gene & Degree & Gene & Degree & Change \\ 
\hline
\emph{TRIM16L} &  24 & \emph{TRIM16L} &  27 &   3 & \emph{TP53BP2} &  22 & \emph{TP53BP2} &  26 &   4 \\ 
  \emph{TP53BP2} &  22 & \emph{CHRNA5} &  25 &   7 & \emph{CDKN2AIPNL} &  21 & \emph{ZMAT3} &  24 &   4 \\ 
  \emph{CDKN2AIPNL} &  19 & \emph{TP53BP2} &  24 &   2 & \emph{CCBE1} &  20 & \emph{CCBE1} &  23 &   3 \\ 
  \emph{PPID} &  19 & \textbf{\emph{CREB1}} &  20 &   2 & \emph{TRIM16L} &  20 & \emph{CDKN2AIPNL} &  23 &   2 \\ 
  \emph{CHRNA5} &  18 & \emph{CDKN2AIPNL} &  19 &   0 & \emph{ZMAT3} &  20 & \emph{KCNH6} &  21 &   5 \\ 
  \textbf{\emph{CREB1}} &  18 & \emph{PPID}$^*$ &  19 &   0 & \emph{BLZF1} &  19 & \emph{TRIM16L} &  20 &   0 \\ 
  \emph{CCBE1} &  17 & \emph{CCBE1} &  18 &   1 & \emph{KIAA0101} &  19 & \emph{ZNF266} &  20 &   4 \\ 
  \emph{KCNH6} &  17 & \emph{ZNF394} &  18 &   4 & \textbf{\emph{CREB1}} &  18 & \emph{BLZF1}$^*$ &  19 &   0 \\ 
  \emph{ZMAT3} &  17 & \emph{KCNH6} &  17 &   0 & \emph{CHRNA5} &  17 & \emph{KIAA0101} &  19 &   0 \\ 
  \emph{FKBP14} &  16 & \emph{KIAA0101} &  17 &   3 & \emph{MCF2L2} &  17 & \emph{MCF2L2} &  19 &   2 \\ 
  \emph{ZNF682} &  16 & \emph{FKBP14}$^*$ &  16 &   0 & \emph{NDUFV3} &  17 & \textbf{\emph{CREB1}} &  18 &   0 \\ 
  \emph{DDX51} &  14 & \emph{ZMAT3} &  16 &  -1 & \emph{PPID} &  17 & \emph{SNRNP48} &  18 &   1 \\ 
  \emph{KIAA0101} &  14 & \emph{ZNF682}$^*$ &  16 &   0 & \emph{SNRNP48} &  17 & \emph{CHRNA5} &  17 &   0 \\ 
  \emph{USP49} &  14 & \emph{AIRE} &  15 &   2 & \emph{KCNH6} &  16 & \emph{NDUFV3} &  17 &   0 \\ 
  \emph{ZNF394} &  14 &  &  &  & \emph{ZNF266} &  16 &  &  &  \\ 
  \emph{ZNF738} &  14 &  &  &  &  &  &  &  &  \\ 
\hline
\end{tabular}
}
\caption{\footnotesize \normalfont Genes with node degrees larger than the 90th percentile in the GM$^*$ and GMSS networks, with the difference in degree between the two networks, for unstimulated (left) and stimulated (right) monocyte analyses. The genes not controlled by the active auxiliary variables are marked with $^*$ in the GMSS network. The genes suspected to be genetic mediators \citep{ruffieux2020global, ruffieux2021epispot} are highlighted in bold.}
\label{tab:hubgenes}
\end{table}

\begin{table}[ht]
\centering
\resizebox{\columnwidth}{!}{%
\begin{tabular}{ccccc|ccccc|cccccc}
\hline
\multicolumn{5}{c|}{\emph{LYZ}} & \multicolumn{5}{c|}{\emph{YEATS4}} & \multicolumn{5}{c}{\emph{CREB1}} \\
\multicolumn{2}{c}{GM$^*$} & \multicolumn{3}{c|}{GMSS} & \multicolumn{2}{c}{GM$^*$} & \multicolumn{3}{c|}{GMSS} & \multicolumn{2}{c}{GM$^*$} & \multicolumn{3}{c}{GMSS}\\
Gene & Degree  & Gene & Degree & Change  & Gene & Degree  & Gene & Degree & Change  & Gene & Degree & Gene & Degree & Change  \\ 
\hline
\emph{TRIM16L} &  24 & \emph{TRIM16L} &  27 &   3 & \emph{TP53BP2} &  22 & \emph{TP53BP2} &  24 &   2 & \emph{TRIM16L} &  24 & \emph{TRIM16L} &  27 &   3 \\ 
  \emph{TP53BP2} &  22 & \emph{CHRNA5} &  25 &   7 & \emph{PPID} &  19 & \emph{KCNH6} &  17 &   0 & \emph{CDKN2AIPNL} &  19 & \emph{CHRNA5} &  25 &   7 \\ 
  \textbf{\emph{CREB1}} &  18 & \emph{TP53BP2} &  24 &   2 & \emph{KCNH6} &  17 & \emph{USP49} &  13 &  -1 & \emph{CHRNA5} &  18 & \emph{CDKN2AIPNL} &  19 &   0 \\ 
  \emph{KCNH6} &  17 &   \textbf{\emph{CREB1}} &  20 &   2 & \emph{USP49} &  14 & \emph{TAF15} &   6 &   0 & \emph{CCBE1} &  17 & \emph{CCBE1} &  18 &   1 \\ 
  \emph{KIAA0101} &  14 & \emph{KCNH6} &  17 &   0 & \emph{TAF15} &   6 &  &  &  & \emph{KCNH6} &  17 & \emph{ZNF394} &  18 &   4 \\ 
  \emph{SEMA3E} &  13 & \emph{KIAA0101} &  17 &   3 &  &  &  &  &  & \emph{ZMAT3} &  17 & \emph{KCNH6} &  17 &   0 \\ 
  \emph{ZNF69} &   9 & \emph{SEMA3E} &  14 &   1 &  &  &  &  &  & \emph{FKBP14} &  16 & \emph{KIAA0101} &  17 &   3 \\ 
  \emph{NDUFV3} &   7 & \emph{ZNF69}$^*$ &   9 &   0 &  &  &  &  &  & \emph{DDX51} &  14 & \emph{FKBP14}$^*$ &  16 &   0 \\ 
   &  & \emph{NDUFV3}$^*$ &   7 &   0 &  &  &  &  &  & \emph{KIAA0101} &  14 & \emph{ZMAT3} &  16 &  -1 \\ 
   &  &  &  &  &  &  &  &  &  & \emph{ZNF394} &  14 & \emph{AIRE} &  15 &   2 \\ 
   &  &  &  &  &  &  &  &  &  & \emph{AIRE} &  13 & \emph{SEMA3E} &  14 &   1 \\ 
   &  &  &  &  &  &  &  &  &  & \emph{SEMA3E} &  13 & \emph{BLZF1}$^*$ &  13 &   1 \\ 
   &  &  &  &  &  &  &  &  &  & \emph{BLZF1} &  12 & \emph{LOC729603} &  13 &   5 \\ 
   &  &  &  &  &  &  &  &  &  & \emph{LRRFIP1} &  12 & \emph{DDX51}$^*$ &  11 &  -3 \\ 
   &  &  &  &  &  &  &  &  &  &   \textbf{\emph{LYZ}} &   9 &   \textbf{\emph{LYZ}$^*$} &  10 &   1 \\ 
   &  &  &  &  &  &  &  &  &  & \emph{ZNF430} &   8 & \emph{LRRFIP1}$^*$ &   9 &  -3 \\ 
   &  &  &  &  &  &  &  &  &  & \emph{RAB27A} &   4 & \emph{ZNF430} &   8 &   0 \\ 
   &  &  &  &  &  &  &  &  &  &  &  & \emph{ZNF786}$^*$ &   6 &   1 \\ 
   &  &  &  &  &  &  &  &  &  &  &  & \emph{RAB27A} &   5 &   1 \\ 
\hline
\end{tabular}
}
\caption{\footnotesize \normalfont Neighbours of \emph{LYZ}, \emph{YEATS4} and \emph{CREB1} ranked by their degrees as estimated by GM$^*$ and GMSS, with difference in degree between the two \textit{unstimulated} networks. \emph{LYZ}, \emph{YEATS4} and \emph{CREB1} are highlighted in bold.
The genes not controlled by active auxiliary variables at 20\% FDR are marked with $^*$ for the GMSS network.}
\label{tab:literaturegenes}
\end{table}

\begin{table}[ht]
\centering
\resizebox{\columnwidth}{!}{%
\begin{tabular}{ccccc|ccccc|cccccc}
\hline
\multicolumn{5}{c|}{\emph{LYZ}} & \multicolumn{5}{c|}{\emph{YEATS4}} & \multicolumn{5}{c}{\emph{CREB1}} \\
\multicolumn{2}{c}{GM$^*$} & \multicolumn{3}{c|}{GMSS} & \multicolumn{2}{c}{GM$^*$} & \multicolumn{3}{c|}{GMSS} & \multicolumn{2}{c}{GM$^*$} & \multicolumn{3}{c}{GMSS}\\
Gene & Degree  & Gene & Degree & Change  & Gene & Degree  & Gene & Degree & Change  & Gene & Degree & Gene & Degree & Change  \\ 
\hline
\emph{TP53BP2} &  22 & \emph{TP53BP2} &  26 &   4 & \emph{TP53BP2} &  22 & \emph{TP53BP2} &  26 &   4 & \emph{CDKN2AIPNL} &  21 & \emph{CCBE1} &  23 &   3 \\ 
  \emph{ZMAT3} &  20 & \emph{ZMAT3} &  24 &   4 & \emph{MCF2L2} &  17 & \emph{MCF2L2} &  19 &   2 & \emph{CCBE1} &  20 & \emph{CDKN2AIPNL} &  23 &   2 \\ 
  \textbf{\emph{CREB1}} &  18 & \emph{KCNH6} &  21 &   5 & \emph{TAF15} &   6 & \emph{TAF15} &   6 &   0 & \emph{TRIM16L} &  20 & \emph{TRIM16L} &  20 &   0 \\ 
  \emph{CHRNA5} &  17 & \textbf{\emph{CREB1}} &  18 &   0 &  &  &  &  &  & \emph{BLZF1} &  19 & \emph{BLZF1}$^*$ &  19 &   0 \\ 
  \emph{NDUFV3} &  17 & \emph{CHRNA5} &  17 &   0 &  &  &  &  &  & \emph{KIAA0101} &  19 & \emph{KIAA0101} &  19 &   0 \\ 
  \emph{SEMA3E} &  14 & \emph{NDUFV3} &  17 &   0 &  &  &  &  &  & \emph{CHRNA5} &  17 & \emph{CHRNA5} &  17 &   0 \\ 
  \emph{TRIM34} &  12 & \emph{TRIM34} &  16 &   4 &  &  &  &  &  & \emph{NDUFV3} &  17 & \emph{NDUFV3} &  17 &   0 \\ 
  \emph{TMEM106A} &   9 & \emph{SEMA3E} &  14 &   0 &  &  &  &  &  & \emph{SEMA3E} &  14 & \emph{TRIM34} &  16 &   4 \\ 
  \emph{TNFSF14} &   6 & \emph{C19orf12} &   9 &   1 &  &  &  &  &  & \emph{LRRFIP1} &  13 & \emph{LOC729603} &  15 &   3 \\ 
  \emph{ZNF131} &   2 & \emph{TMEM106A} &   8 &  -1 &  &  &  &  &  & \emph{LOC729603} &  12 & \emph{QRFPR} &  14 &   2 \\ 
   &  & \emph{TNFSF14} &   6 &   0 &  &  &  &  &  & \emph{QRFPR} &  12 & \emph{SEMA3E} &  14 &   0 \\ 
   &  & \emph{ZNF131} &   3 &   1 &  &  &  &  &  & \emph{TRIM34} &  12 & \emph{BMS1P5} &  13 &   2 \\ 
   &  &  &  &  &  &  &  &  &  & \emph{BMS1P5} &  11 & \textbf{\emph{LYZ}$^*$} &  13 &   2 \\ 
   &  &  &  &  &  &  &  &  &  & \textbf{\emph{LYZ}} &  11 & \emph{LRRFIP1} &  12 &  -1 \\ 
   &  &  &  &  &  &  &  &  &  & \emph{USP49} &  10 & \emph{USP49} &  10 &   0 \\ 
   &  &  &  &  &  &  &  &  &  & \emph{TMEM106A} &   9 & \emph{TMEM106A} &   8 &  -1 \\ 
   &  &  &  &  &  &  &  &  &  & \emph{HNRNPU} &   4 & \emph{PPM1K}$^*$ &   4 &   1 \\ 
\hline
\end{tabular}
}
\caption{\footnotesize \normalfont Neighbours of \emph{LYZ}, \emph{YEATS4} and \emph{CREB1} ranked by their degrees as estimated by GM$^*$ and GMSS, with difference in degree between the two \textit{stimulated} networks. \emph{LYZ}, \emph{YEATS4} and \emph{CREB1} are highlighted in bold.
The genes not controlled by active auxiliary variables at 20\% FDR are marked with $^*$ for the GMSS network.}
\label{tab:literaturegenes_stimulated}
\end{table}
\end{appendices}
\end{document}